%% file: main.tex
\documentclass[%
 reprint,
 amsmath,amssymb,
 aps,
prb,
]{revtex4-2}

\usepackage{graphicx}% Include figure files
\usepackage{dcolumn}% Align table columns on decimal point
\usepackage{bm}% bold math
\usepackage{braket}%braket symbols
\usepackage{xcolor}
\usepackage[normalem]{ulem}
\usepackage[abs]{overpic}

\newcommand{\be}{\begin{eqnarray}}
\newcommand{\ee}{\end{eqnarray}}
\newcommand{\nn}{\nonumber}
\newcommand{\bea}{\begin{eqnarray}}
\newcommand{\eea}{\end{eqnarray}}

\renewcommand{\vec}[1]{{\bm #1}}

\begin{document}

\preprint{APS/123-QED}

\title{Spin and Charge Transport Mediated by Fractionalized Excitations}% Force line breaks with \\
%\thanks{A footnote to the article title}%

\author{Alexey Ermakov}
 \email{Alexey.Ermakov@manchester.ac.uk}
\author{Alessandro Principi}%
 \email{Alessandro.Principi@manchester.ac.uk}
\affiliation{%
 Department of Physics and Astronomy, University of Manchester, Oxford Road, M13 9PL Manchester, United Kingdom
}%

\date{\today}% It is always \today, today,
             %  but any date may be explicitly specified

\begin{abstract}
We study charge and spin transport in systems composed of itinerant electrons and localized magnetic moments of a Kitaev quantum spin liquid (QSL) phase. For example, $\alpha-{\rm RuCl}_3$ either intrinsically doped or in proximity to graphene. 
Our analysis reveals distinct temperature-dependent transport behaviors due to the QSL gap and the Majorana excitation spectrum, which are greatly enhanced when the velocities of itinerant electrons and QSL excitations become comparable. These transport characteristics could potentially be employed to reveal the fractionalized excitations of the spin liquid.
\end{abstract}

%\keywords{Suggested keywords}%Use showkeys class option if keyword
                              %display desired
\maketitle

%\tableofcontents

\section{\label{sec:intro} Introduction}
Quantum spin liquids (QSL) can emerge as ground states of frustrated quantum spin systems, where competing magnetic interactions prevent the establishment of long-range order even at zero temperature~\cite{zhou2017quantum, savary2016quantum, wen2019experimental}. Instead, the system adopts a state characterized by long-range entanglement and fractionalized (possibly non-Abelian) excitations interacting with emergent gauge fields~\cite{knolle2019field, wen2004quantum}. These properties make QSLs particularly intriguing for both theoretical investigations and potential applications in quantum computing and information processing~\cite{nayak2008non,stern2008anyons,blasi2012non}. Despite their theoretical appeal, experimental realizations of QSLs remain scarce \cite{muller2022one,takagi2019concept}. The sensitivity of QSL states to microscopic parameters means that moderate variations in material composition or external conditions can drive the system out of the QSL phase, complicating the search for materials that exhibit this exotic behavior.

Recent advancements have identified $\alpha$-RuCl$_3$ as a promising candidate to realize a Kitaev honeycomb model~\cite{takagi2019concept,loidl2021proximate,yadav2016kitaev}.
This model~\cite{kitaev2006anyons} features bond-dependent Ising interactions on a two-dimensional honeycomb lattice. Due to its exactly solvable nature, the model's ground and excited states can be exactly characterized. In particular, the highly entangled ground state features fractionalized excitations such as itinerant Majorana fermions and localized visons~\cite{hermanns2018physics}. The exotic excitations of the Kitaev model have significant implications for quantum computation, as they provide a platform for the creation of topologically protected quantum bits. 
% Fractionalization of a spin degree of freedom leads to emergence of itinerant spinon (mobile excitations) and three localized bond gauge fermions. 
Theoretical studies have also shown that the Kitaev model supports a rich phase diagram with various phases depending on the relative strengths of the bond interactions, external perturbations~\cite{knolle2016dynamics} and doping~\cite{peng2021precursor}, some of which are believed to be QSLs of a different kind.

Kitaev interactions in $\alpha$-RuCl$_3$ are generated by the complex interplay between strong spin-orbit coupling, crystal field, and interference between super-exchange pathways~\cite{PhysRevLett.102.017205}.
% , making it conducive to Kitaev interactions. 
Experimental observations have shown that under applied magnetic fields and pressure, $\alpha$-RuCl$_3$ displays signatures consistent with a Kitaev QSL~\cite{baek2017evidence}. 
The van-der-Waals nature of this material
allows to consider the interaction with an
electronic system in its proximity, such as a graphene layer. 

When placed directly in contact with graphene, charges are transferred into $\alpha$-RuCl$_3$. The charges injected in the material interact with localized spins via a local spin-spin coupling. This enables a complex interplay between charge and spin transport mediated by the carriers injected in the material.
Even when no charge transfer is allowed between graphene and $\alpha$-RuCl$_3$, the transport of charge and spin are influenced by the interaction between the spin of itinerant electrons and the localized magnetic moments. This unique characteristic could make
quantum transport a promising route for on-chip probing of exotic QSL excitations, thus improving our understanding of this complex quantum state and its potential applications.

Motivated by this unique possibility offered by $\alpha$-RuCl$_3$/graphene heterostructures, in this paper we derive a kinetic equation for electrons interacting with QSL spins and 
% apply linear response theory to 
determine the electric and spin responses of such systems. We assume that QSL spinons and free electrons interact via a local Kondo-like coupling between their spin degrees of freedom. We assume a linear dispersion relation for the electrons, having in mind applications to graphene-based van-der-Waals heterostructures. We find that, when electrons are not spin polarized, both the transport collision rate and spin diffusion rate exhibit an activation behavior consistent with spin excitations being gapped. Both rates increase approximately linearly with temperature $T$, and tend to saturate when $k_{\rm B} T$ ($k_{\rm B}$ is Boltzmann's constant) becomes larger than the bandwidth. When the electron's Fermi velocity is reduced, both collision rate markedly increase. When electrons are partially spin-polarized, the transport time remains nearly independent of spin polarization. Conversely, the spin-flip collision rate becomes finite for any infinitesimal spin polarization.

The paper is structured as follows: Section II provides a theoretical background of the Kitaev model and the derivation of the kinetic equation. Section III outlines the derivation of the spin structure factor, which is the main physical quantity entering the collision integral. Section IV discusses the results and implications of our findings, with an emphasis on the potential observable effects in experimental setups involving $\alpha$-RuCl$_3$ and graphene heterostructures.

\section{Kinetic equation and collision integral}
We consider electrons interacting with the localized moments of a Kitaev system via a Kondo-like coupling. The Hamiltonian is ${\cal H}={\cal H}_0+{\cal H}_K+V$, where ${\cal H}_0$ is the electron Hamiltonian
\begin{equation}
    {\cal H}_0 = \sum_{\bm k, \sigma} \varepsilon_{\bm k, \sigma} c^{\dagger}_{\bm k, \sigma} c_{\bm k, \sigma}.
\end{equation}
Here, $c^{\dagger}_{\bm k, \sigma} (c_{\bm k, \sigma})$ creates (annihilates) an electron of momentum $\bm k$ and spin projection $\sigma=\{ \uparrow, \downarrow \}$, whose energy is $\varepsilon_{\bm k, \sigma}$. The creation and annihilation operators satisfy the usual anti-commutation relations $\{  c_{\bm k,\sigma}, c_{\bm p,\sigma '}^{\dagger} \} = \delta_{\bm k, \bm p}\delta_{\sigma, \sigma '}$. 

The Kitaev Hamiltonian describing the dynamics of spin-$1/2$ quantum magnetic moments of the QSL is 
\begin{equation} \label{eq:H_K_def}
    {\cal H}_K = -J_K\sum_{\braket{i,j}_\gamma} \bm s^\gamma_{i} \bm s^\gamma_j ,
\end{equation}
where $\bm{s}_i = (s^x_{i},s^y_{i},s^z_{i})$ are spin operators at the sites of a honeycomb lattice and $J_K$ is the Kitaev bond-dependent Ising-like interaction between nearest neighbors,
% in principle depends on the bond direction. 
the sum runs over nearest neighbors $i$ and $j$ in direction $\gamma$ and $\gamma \in \{ x,y,z \}$ denotes the three bond orientations. For a sake of simplicity, in Eq.~(\ref{eq:H_K_def}) we have taken the case of isotropic interactions. 
This model is believed to be a good approximation for the physics of RuCl$_3$ at temperatures above $\approx 7$ K, 
% {\it i.e.} above the ordering temperature
below which the material exhibits conventional order~\cite{loidl2021proximate}. 
We use the value of the coupling constant for RuCl$_3$ to be
$J_K\approx 1.8$ meV, which is within a wide range of estimates~\cite{carrega2020tunneling, PhysRevLett.114.147201,hermanns2018physics}.

Finally, the Kondo-like interaction is
\begin{equation}
    V=J\sum_i \bm{s}_i^e\cdot \bm{s}_i,
\end{equation}
where $J$ is the coupling constant {arising from weak van der Waals interlayer exchange \cite{Zhou2019_PRB_graphene_RuCl3, Mashhadi2019_NanoLett_Graphene_RuCl3}} and  $\bm{s}_i^{e}$ is electron spin operator at site $i$:
\begin{equation}
    \bm s_i^e = c^{\dagger}_{i, \sigma} \bm \sigma_{\sigma \sigma'} c_{i, \sigma'},
\end{equation}
where $\bm \sigma$ is a vector of Pauli matrices. {We work perturbatively in Kondo coupling $J$. The dimensionless parameter governing the perturbative expansion is $\alpha = \rho(\mu)J$ where $\rho = |\mu|/[2\pi (v_0 \hbar)^2]$ is the Dirac density of states per spin and valley, $\mu$ is the chemical potential, $v_0$ is Fermi velocity of itinerant electrons ($10^6$ m/s in graphene). For $\alpha$-RuCl$_3$-graphene heterostructures $\alpha\sim 10^{-4}-10^{-5}$ for coupling $J\sim 2-10$ meV \cite{PhysRevLett.126.097201}.  } \\

The annihilation operator in real space is constructed from $c_{\bm k, \sigma}$ as 
\begin{equation}
    c_{i,\sigma} = \frac{1}{\sqrt{N}}\sum_{\bm k}c_{\bm k,\sigma} e^{-i\bm k \cdot \bm R_i}.
\end{equation}
Here, $\bm R_i$ is a vector in real space denoting the location of the lattice site $i$, while $N$ is total number of lattice sites. With this definition, the interaction Hamiltonian can be rewritten
\begin{equation}
    V = J\sum_{\bm k, \bm k', \sigma, \sigma'} c_{{\bm k}^{\color{red}{'}},\sigma}^{\dagger} (\bm \sigma_{\sigma \sigma'} \cdot \bm s^K_{\bm k-\bm k'})  c_{\bm k,\sigma}.
\end{equation}

The transport properties of the system subject to the electric field $\bm E$, can be found by solving the Boltzmann equation
\begin{equation}
    (\partial_t + \bm v_{\bm k}\cdot \bm \nabla_{\bm r} - e \bm E\cdot \bm \nabla_{\bm k} ) f_{\bm k, \sigma} = (\partial_tf_{\bm k, \sigma})_{\text{coll}}.
\end{equation}
To derive this, we need to consider the time evolution of electron distribution function, which is given by
\begin{equation}
    f_{\bm k, \sigma} = \braket{c_{\bm k,\sigma}^{\dagger}(t)c_{\bm k,\sigma}(t)}.
\end{equation}
Here, the average is taken over the canonical ensemble described by the density matrix $e^{-\beta {\cal \hat{H}} }$, {\it i.e.} $\braket{\hat{A}} = \mathrm{Tr}[e^{-\beta {\cal \hat{H}} }\hat{A}]$ where $\beta = 1/k_B T$ is the inverse temperature. The collision integral can be found from the time derivative of $ f_{\bm k, \sigma}$ in the absence of driving fields, {\it i.e.}
\begin{equation}
    (\partial_tf_{\bm k, \sigma})_{\text{coll}} = -i\braket{[{\cal H}, c_{\bm k,\sigma}^{\dagger}(t)c_{\bm k,\sigma}(t)]}.
\end{equation}
To proceed we use perturbation theory to the lowest nonvanishing order, {\it i.e.} to second order in $V$. In this case, the time evolution of creation and annihilation operators is governed by the free Hamiltonian ${\cal H}_0$: 
\begin{equation}
    c_{\bm k,\sigma}(t) = c_{\bm k,\sigma}e^{-i\varepsilon_{\bm k, \sigma}t},
\end{equation}
and the collision integral is 
\begin{equation} \label{eq:coll_integral_second_order}
    (\partial_tf_{\bm k, \sigma})_{\text{coll}} = \int_{-\infty}^t dt' \braket{[V(t),[V(t'), c_{\bm k,\sigma}^{\dagger}(t)c_{\bm k,\sigma}(t)]]},
\end{equation}
where $V(t) = {{e^{i{\cal H}_0 t} V e^{-i{\cal H}_0 t}}}$ is the Kondo-like Hamiltonian in the interaction picture. 

Since we consider only the lowest nonvanishing order in the Kondo coupling, the average $\braket{\ldots}$ in Eq.~(\ref{eq:coll_integral_second_order}) is taken over the non-interacting density matrix.
This factorizes into a product of the electrons and quantum spin liquid density matrices. Therefore, the expectation value can also be factorized as
\begin{eqnarray}
    &&\braket{c_{\bm q,\alpha}^{\dagger} c_{\bm q',\alpha'} c_{\bm p,\lambda}^{\dagger} c_{\bm k,\sigma} s^\eta_{\bm q-\bm q'}(t)s^\gamma_{\bm p -\bm k}(t')}= \nonumber \\
    &&\braket{c_{\bm q,\alpha}^{\dagger} c_{\bm q',\alpha'} c_{\bm p,\lambda}^{\dagger} c_{\bm k,\sigma}} \braket{s^\eta_{\bm q-\bm q'}(t)s^\gamma_{\bm p -\bm k}(t')}.
\end{eqnarray}
%
%where the spin operator $s^\eta_{\bm q}$ is the $\eta^{\text{th}}$ component of the vector $\bm s^K_{\bm q}$, $\eta = \{ x,y,z \}$. 
The details of the calculation are given in App.~\ref{app:collintder}. Here, we only present the final result, which reads
\begin{widetext}
\begin{equation}\label{coll_int}
    (\partial_tf_{\bm k, \sigma})_{\text{coll}} = 
    \frac{J^2}{\pi} \sum_{\bm p,\lambda, \gamma} \int \limits_{-\infty}^\infty d \omega \delta\left( \varepsilon_{\bm k,\sigma}-\varepsilon_{\bm p, \lambda}-\omega \right) \sigma^\gamma_{\sigma \lambda}\sigma^\gamma_{\lambda \sigma} \left[ f_{\bm k, \sigma}(1-f_{\bm p, \lambda})\langle s^\gamma_{\bm k - \bm p}(\omega) s^\gamma_{\bm p - \bm k} \rangle - f_{\bm p, \lambda}(1-f_{\bm k, \sigma}) \langle s^\gamma_{\bm p - \bm k}(-\omega) s^\gamma_{\bm k - \bm p} \rangle \right].
\end{equation}
\end{widetext}
This collision integral depends on the spin structure factor ${ F }(\vec{q},\omega) = \sum_{\gamma}\langle s^\gamma_{\bm q}(\omega) s^\gamma_{-\bm q} \rangle$ which contains all the information about the spin excitation of the quantum spin liquid. The next section describes its derivation at finite temperature, correcting some mistake existing in the literature.

\section{The spin structure factor}
Before showing the calculation of the spin structure factor we briefly review the exact solution of the Kitaev model, which passes through the fractionalization of spins into fermionic excitations, as we discuss below. 

\subsection{Kitaev model of $\alpha$-R\lowercase{u}C\lowercase{l}$_3$}

Following Kitaev~\cite{kitaev2006anyons}, we introduce the four Majorana fermions $c_i$ and $b_i^\gamma$ ($\gamma=x,y,z$) at each lattice site $i$, such that 
\begin{equation} \label{fractionalization}
    s_i^\gamma = i b_i^\gamma c_i .
\end{equation}
This Majorana operators satisfy the anticommutation relations $\{ c_i,c_j \} = 2 \delta_{i,j}$, $\{ b_i^\gamma , b_j^\eta \} = 2 \delta_{i,j} \delta^{\gamma, \eta}$ and $\{ c_i , b_j^\eta \} = 0$. Written in terms of Majorana operators, the Kitaev Hamiltonian becomes 
\begin{equation}\label{Hk_with_U}
    {\cal H}_K = iJ\sum_{\braket{i,j}_\gamma} u_{ij}^\gamma c_i c_j.
\end{equation}
All $u_{ij}^\gamma = i b_i^\gamma b_j^\gamma$ commute with the Hamiltonian and are therefore constants of motion. 
%Since ${\cal H}_K$ and $u_{ij}^\gamma$ share the same set of eigenstates, 
Working with common eigenstates of ${\cal H}_K$ and $u_{ij}^\gamma$, 
we can replace all operators $u_{ij}^\gamma$ with their eigenvalues, which are equal to $\pm 1$ by construction.
The map \eqref{fractionalization} artificially enlarges the Hilbert space of the theory. This redundancy can be taken care of by projecting the states on the physical subspace. The projection operator is defined as $D_j = -i s_j^x s_j^y s_j^z = b_j^x b_j^y b_j^z c_j $ and physical states satisfy $D_j\ket{\text{phys}} =  \ket{\text{phys}}$ for all $j$. 

The bond operators $u_{ij}^\gamma$ play a role of an effective ``magnetic" $\mathbb{Z}_2$ vector potential. The potential generates $\mathbb{Z}_2$ ``magnetic fluxes" piercing through plaquettes: $\Phi_p = \prod_{\braket{i,j}^\gamma \in p}u_{ij}^\gamma$, where ``$\braket{i,j}^\gamma \in p$'' denotes all bonds belonging to the plaquette $p$. Therefore, once we substitute the eigenvalues of $u_{ij}^\gamma$ into \eqref{Hk_with_U} we get a Hamiltonian for free fermionic particles, $c_i$, propagating on top of magnetic fluxes threading the honeycomb plaquettes. 
The ground state is found in the portion of the Hilbert space corresponding to setting all $u_{ij}^\gamma = -1$, effectively leaving no magnetic fluxes in the system. 

\subsection{Correlation function}
The spin structure factor $F(\vec{q},\omega)$ is related to the imaginary part of the spin-spin response function $\chi (\vec{q},\omega)$ via the fluctuation-dissipation theorem: 
\begin{equation}
    F(\vec{q},\omega) = (n_B(\omega)+1) \sum_{\gamma} \Im m\chi^{\gamma} (\vec{q},\omega),
\end{equation}
where $n_B(\omega)=[ \exp(\beta \omega)-1 ]^{-1}$ is the Bose-Einstein distribution function and $\beta = (k_BT)^{-1}$. 
Therefore, the problem of calculating $F(\vec{q},\omega)$ is reduced to calculating the imaginary-time-ordered function 
\begin{equation} \label{response}
    \chi_{ij}^\gamma(\tau) = -\braket{\text{T} s_i^\gamma (\tau) s_j^\gamma} = \text{tr} \left[ e^{-\beta {\cal H}_K} \text{T} s_i^\gamma (\tau) s_j^\gamma \right] .
\end{equation}

Here we only summarize the calculation of this function and refer the reader to App.~\ref{app:bosesum} for more details. The idea is to apply the Kitaev's mapping \eqref{fractionalization} to the response function \eqref{response} and evaluate the latter on the ground-state flux sector. The problem maps into that of finding the Green's function of the $c$ fermions in the presence of a localized impurity
\begin{equation}\label{eq:GF1}
    \chi_{ij}^\gamma (\tau) = \text{tr}\left[ e^{-\beta {\cal H}_K} {\rm T} c_i(\tau)c_j S_{i,\gamma}(\tau, 0) \right], 
\end{equation}
where the impurity scattering matrix is
\begin{equation} \label{eq:s-matrix}
    S_{i,\gamma}(\tau_1, \tau_2) = {\rm T} \exp \left( 2 i J_K \int \limits_{\tau_1}^{\tau_2} d\tau' \sum_{\braket{i,l}_\gamma} c_i(\tau') c_l (\tau')  \right) .
\end{equation}
To proceed we use the so-called adiabatic approximation~\cite{carrega2020tunneling}, rewriting \eqref{eq:GF1} as 
\begin{equation} \label{eq:QSL_adiabatic_approx}
    \chi_{ij}^\gamma(\tau) \simeq \frac{ \braket{{\rm T} c_i(\tau)c_j S_{i, \gamma}(\beta,0)} }{ \braket{{\rm T} S_{i, \gamma}(\beta,0)} } \braket{{\rm T} S_{i, \gamma}(\tau,0)}.
\end{equation}
The factor $ G_{ij}^{\gamma,c}(\tau)=\braket{{\rm T} c_i(\tau)c_j S_{i, \gamma}(\beta,0)}/ \braket{{\rm T} S_{i, \gamma}(\beta,0)} $ can be calculated exactly by resumming a series of connected Feynman diagrams and is in fact the connected Green's function of the $c$ fermions in the presence of the impurity potential. \\

{
The question of applicability of the adiabatic approximation, Eq.~(\ref{eq:QSL_adiabatic_approx}), has been discussed in \cite{knolle2016dynamics}. Since the calculation is effectively mapped onto a single-impurity problem, the main question is whether 
% the adiabatic approximation leads to 
an orthogonality catastrophe~\cite{Anderson} occurs. This turns out not to be the case since the Kitaev QSL dispersion is linear at the neutrality point~\cite{kitaev2006anyons}. 
% Therefore, 
The orthogonality catastrophe is avoided
because the density of state vanishes at the neutrality point.
% does not occur. 
In~\cite{knolle2016dynamics} it was shown that the overlap between the 
% true 
ground state 
in the presence of a flux pair,
$\ket{M_{\rm F}}$, and the unperturbed one, $\ket{M_{\rm 0}}$, is $|\braket{M_{\rm F}|M_{\rm 0}}|\approx 0.77$ in the thermodynamic limit. This indicates that the adiabatic approximation is sufficiently accurate to capture the main features of the problem.
}
The second factor in Eq.~\eqref{eq:QSL_adiabatic_approx} can be approximated as~\cite{carrega2020tunneling} 
\begin{equation}\label{approxS}
    \braket{{\rm T} S_{i, \gamma}(\tau,0)} \simeq e^{-\Delta \tau}, 
\end{equation}
where $\Delta$ is the minimum energy of spin excitations \cite{kitaev2006anyons, knolle2016dynamics}. 
{To interpret our results, we outline the energy scale hierarchy. The Kondo coupling is the smallest scale at $J \sim 1$ meV, while the Kitaev coupling $J_K \approx 1.8-5$ meV (with vison gap $\Delta = 0.26 J_K \approx 0.468$ meV). The electron chemical potential $\mu$ is tunable between 0-200 meV via doping, while the bandwidth of the linear part of electrons' dispersion in graphene is the largest scale and equals $W \sim 2-3$ eV \cite{knolle2016dynamics,Joy2022_PRX_VisonDynamics}. 
The temperature $k_BT$ ranges from sub-meV to $\sim$25 meV ($\sim$300 K), spanning the range from below the QSL gap and up to the part of bandwidth $W$. 
Since $J \ll J_K$, perturbation theory is applicable. The order of energy scales also explains features like the activation threshold in Figs.~\ref{fig:spindeg300}-\ref{taudiff} at $T \sim \Delta$ ($\sim 5$ K), {\it i.e.} the onset of flux excitations, and the linear-in-T growth above the QSL bandwidth ($\sim J_K \sim20$ K). For lower $v_0$ (e.g., doped $\alpha$-RuCl$_3$ bands), all rates scale as $v_0^{-2}$, enhancing the impact of the electron-spin coupling (as in Fig.~\ref{fig:spindeg300}(b)).}

However, one has to be careful about the overall statistics of the response function \eqref{response}. Notice, that \eqref{response} carries bosonic statistics as it is made of two commuting operators. Meanwhile, the connected Green's function $G_{ij}^c(\tau)$ is fermionic and the approximated $S$-matrix \eqref{approxS} has no well-defined statistics at all. Yet, we have to preserve the right periodicity of the response function in imaginary time. 

To do so, we assume \eqref{approxS} is true for $\tau>0$ and impose the anti-periodicity condition 
\begin{equation}
    \braket{{\rm T} S_{i, \gamma}(\tau-\beta,0)}=-\braket{{\rm T} S_{i, \gamma}(\tau,0)}, \quad 0 < \tau < \beta,
\end{equation}
so that the overall response function \eqref{response} remains periodic. 
%Here $\beta = 1/T$ is the inverse temperature. 
One of the possible ways to write down the $S$-matrix consistent with the anti-periodicity requirement is 
\begin{eqnarray}
    &&\braket{{\rm T} S_{i, \gamma}(\tau,0)} = \nonumber \\ 
    &&\frac{1+e^{-\beta \Delta}}{\beta}\sum_{\epsilon_k=-\infty}^{\infty}e^{-i\epsilon_k \tau}\left( \frac{\theta(\tau)}{\Delta-i \epsilon_k}+\frac{\theta(-\tau)}{\Delta + i \epsilon_k} \right) ,
\end{eqnarray}
where $\epsilon_k$ are fermionic Matsubara frequencies. 

We combine Majorana $c$-fermions from two sub-lattices to form a complex fermion $f_{\bm r} = (c_{\bm r, A} + i c_{\bm r, B})/2$ and $f_{\bm r}^\dagger = (c_{\bm r, A} - i c_{\bm r, B})/2$. The Green's function for $c$-fermions can then be calulated in terms of connected Green's functions for $f$-fermions 
\begin{equation} \label{eq:Q_c_gamma_fermions}
    Q_c^\gamma(\bm r, \tau, \tau') = - \frac{ \braket{{\rm T} f_{\bm r}(\tau) f^\dagger_{\bm r}(\tau') S_{\bm r, \gamma}(\beta)} }{\braket{ {\rm T} S_{\bm r, \gamma}(\beta)} } .
\end{equation}
{
We note the dependence of $Q_c^\gamma(\bm r, \tau, \tau')$ on bond direction $\gamma$ stemming from the S-matrix definition Eq.~(\ref{eq:s-matrix}) and from the definition of $f$-fermions, which contain implicit information about bond directions as they are constructed with $c$-fermions from different sublattices. 
}

Resummation of all Feynman diagrams gives in frequency space \cite{carrega2020tunneling}
\begin{equation}
    Q_c^\gamma(\bm r, \omega) = \frac{ Q_{c,0}^\gamma (\bm r, \omega) }{ 1 + 4J Q_{c,0}^\gamma (\bm r, \omega) } ,
\end{equation}
where $Q_{c,0}^\gamma (\bm r, \omega)$ is the bare Green's function obtained by setting $S_{\bm r}(\beta) = 1$ in Eq.~(\ref{eq:Q_c_gamma_fermions}). {\it I.e.}, it can be obtained by Fourier transforming $Q_{c,0}^\gamma(\bm r, \tau, \tau') = \braket{{\rm T} f_{\bm r}(\tau) f^\dagger_{\bm r}(\tau')} .$

Next, we calculate $Q_{c,0}^\gamma (\bm r, \omega)$ and see that it actually does not depend on position $\bm r$. Therefore, in what follows, we omit the $\bm r$ dependence from both bare and connected Green's functions. After this, we analytically continue the obtained response function $\chi$ to real frequencies and finally get the imaginary part of the response function in momentum-frequency space
\begin{widetext}
\begin{eqnarray} \label{eq:Imchi_final}
    &&\Im m\chi^\gamma(\omega, \bm{k}) = (1+e^{-\beta \Delta})\nonumber \\
    &&\times \left\{ [(n_F(\omega-\Delta)-n_F(-\Delta)) \Im mQ_c^\gamma(\omega-\Delta) - (n_F(-\omega-\Delta)-n_F(-\Delta)) \Im mQ_c^\gamma(-\omega-\Delta)](1+\cos (\bm k \cdot \bm{d}^\gamma)) \right. \nonumber \\
    && + \left.  [(n_F(\omega-\Delta)-n_F(-\Delta)) \Im mQ_c^\gamma(-\omega+\Delta) - (n_F(-\omega-\Delta)-n_F(-\Delta)) \Im m Q_c^\gamma(\omega+\Delta)](1-\cos (\bm{k} \cdot \bm{d}^\gamma)) \right\} , 
    %\label{imchi}
\end{eqnarray}
\end{widetext}
where $\bm d^\gamma$ are positions of three nearest neighbors of atom of type $A$ in directions $\gamma = \{ x,y,z \}$ and we recall that $\Delta$ is the energy cost of exciting a pair of fluxes in neighboring plaquettes. Equation~(\ref{eq:Imchi_final}) corrects the expression given in Ref.~\cite{carrega2020tunneling} ensuring the correct periodicity properties of its Fourier transform.

The expression \eqref{eq:Imchi_final} is explicitly antisymmetric in $\omega$: $\Im m\chi^\gamma(\omega, \bm{k}) = - \Im m\chi^\gamma(-\omega, \bm{k})$ as expected from the analytic properties of a response function. Note also, that $\Im m\chi^\gamma(\omega, \bm{k}) = \Im m\chi^\gamma(\omega, - \bm{k})$ is even under momentum inversion. These properties will prove useful in what comes next. 

\section{Linear response}
We start with rewriting the collision integral \eqref{coll_int} using all the obtained results so far and expanding the distribution function around its equilibrium value $f_{\bm k, \sigma}=~ f^0_{\bm k, \sigma}+~\delta f_{\bm k, \sigma}$, where $f^0_{\bm k, \sigma}=[1+\exp\big(\beta(\varepsilon_{\bm k, \sigma}-\mu)\big)]^{-1}$ is the Fermi distribution function. Plugging this into \eqref{coll_int} and keeping only terms linear in $\delta f$ we get 
\begin{widetext}
\begin{eqnarray} 
    (\partial_tf_{\bm k, \sigma})^{\rm lin}_{\text{coll}} &=& 
    \frac{J^2}{\pi} \sum_{\bm p,\lambda, \gamma} \int \limits_{-\infty}^\infty d \omega \delta\left( \varepsilon_{\bm k,\sigma}-\varepsilon_{\bm p, \lambda}-\omega \right) 
    \sigma^\gamma_{\sigma \lambda}\sigma^\gamma_{\lambda \sigma} \Im m \chi_{\bm k - \bm p}^\gamma
    (1-f^0_{\bm k, \sigma}) (1-f^0_{\bm p, \lambda})(n_B(\omega)+1) \nonumber \\
    &\times& \left[ \frac{ \delta f_{\bm k, \sigma} }{ (1-f_{\bm k, \sigma})^2 } - \frac{ \delta f_{\bm p, \lambda} }{ (1-f_{\bm p, \lambda})^2} \frac{n_B(\omega)}{n_B(\omega)+1} \right] . \label{coll_int_2}
\end{eqnarray}
\end{widetext}
We assume the system is isotropic. In the steady state, the linearized Boltzmann equation becomes
\begin{equation} \label{bollzeq}
    -e \bm E \cdot \bm v_{\bm k} \frac{ \partial f^0_{\bm k, \sigma} }{\partial \varepsilon_{\bm k, \sigma}} = (\partial_tf_{\bm k, \sigma})_{\text{coll}}^{\rm lin} .
\end{equation}
From here on we can study both electric and spin currents. Since, the case of (partially) spin-polarized electrons is quite complicated, we start with the spin-unpolarized case to build an intuition and crosscheck the results we will obtain for spin current on the more general case.

\begin{figure}[t]
\begin{tabular}{c}
\includegraphics[width = \linewidth]{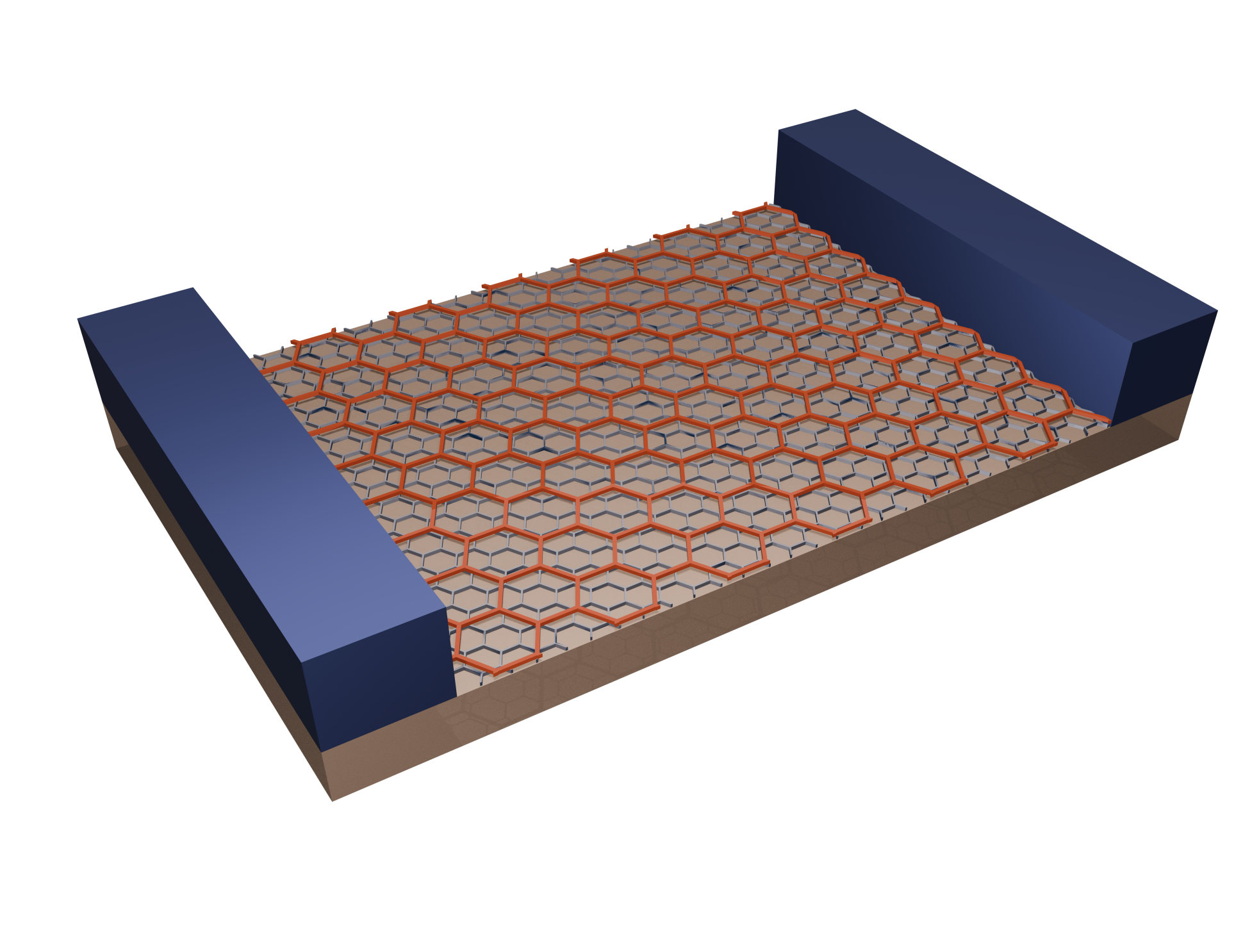}
% fig_setup.pdf}
\end{tabular}
\caption{\label{fig:setup} 
{Schematics of the setup.} A Kitaev QSL material (e.g., $\alpha$-RuCl$_3$) is deposited on a substrate and placed in proximity to a metallic (e.g., graphene) layer. Direct contact can induce charge transfer and doping of the QSL~\cite{Zhou2019_PRB_graphene_RuCl3,Rossi2023_NanoLett_ChargeTransferImaging}. Ferromagnetic and/or high-spin-orbit electrodes can be used to inject and detect charge and spin currents/density accumulations.
}
\end{figure}

\subsection{Spin-unpolarized case I: transport time}

{
Here we calculate the transport time for electrons when they interact with the excitation of the QSL. As any other bosonic quasiparticle (e.g., phonons), spinons provide a source of (electric) current relaxation. The setup we have in mind is schematically shown in Fig.~\ref{fig:setup}: a Kitaev QSL (e.g., an ${\alpha-\rm RuCl}_3$ monolayer) is placed in proximity to a metallic system (e.g., a graphene layer). Proximity can result in doping of the QSL~\cite{Rossi2023_NanoLett_ChargeTransferImaging,Plumb2014_PRB_alphaRuCl3,Sinn2016_SciRep_alphaRuCl3_PE_IPE}. Electric contacts inject currents into either the metallic system or (if doped) into the QSL. Our simplifying assumption is that both systems have the same linear energy dispersion \cite{PhysRevLett.126.097201}, albeit with different Fermi velocities (the latter being much smaller for the QSL due to the flatness of the band structure~\cite{Rossi2023_NanoLett_ChargeTransferImaging,Plumb2014_PRB_alphaRuCl3,Sinn2016_SciRep_alphaRuCl3_PE_IPE}). This assumption is not restrictive as most of the conduction occurs close to the Fermi energy, where bands can be linearized. The transport time can be compared with the measured one. The latter is obtained, knowing the effective mass and charge density, from measurements of the electrical resistivity according to the usual Drude formula.
}

To get the transport we sum over the spin indices in equations \eqref{coll_int_2}-\eqref{bollzeq}. To proceed, we are going to use the relaxation time approximation introducing the transport relaxation time $\tau_{\rm tr}$ as follows 
\begin{equation} \label{eq:unpol_BE_RTA}
    -e \bm E \cdot\bm v_{\bm k} \frac{ \partial f^0_{\bm k} }{\partial \varepsilon_{\bm k} } = -\frac{\delta f_{\bm k}}{\tau_{\rm tr}},
\end{equation}
where we assumed the transport time to be independent of energy. The goal here is to relate $\tau_{\rm tr}$, as a function of temperature and carrier concentration, to the collision integral~\eqref{coll_int_2}.

We parameterize the angle between electric field $\bm E$ and electron's velocity $\bm v_{\bm k} $ as $\phi_{\bm k}$. To determine $\tau_{\rm tr}$ we invert Eq.~\eqref{eq:unpol_BE_RTA} and express the deviation of the distribution function from equilibrium, in terms of the relaxation time 
\begin{equation}
    \delta f_{\bm k} = \tau_{\rm tr} e E v_{\bm k} \cos \phi_{\bm k} \frac{ \partial f^0_{\bm k} }{\partial \varepsilon_{\bm k} }
\end{equation}
and plug it back in the microscopically derived collision integral \eqref{coll_int_2}. We then multiply both sides of the kinetic equation by $v_{\bm k} \cos \phi_{\bm k}$ and integrate over $\bm k$. The details of these calculations, as well as definitions of functions $f(\omega)$ and $g(\omega)$, are given in App.~\ref{app:unpolCR}. The resulting transport collision rate is 
%kinetic equation takes the form 
\begin{eqnarray}
    && \tau_{\rm tr}^{-1}  =  \frac{ J^2}{2 \pi T {\cal D}}  \int d k d p d\omega k p \delta(\varepsilon_{\bm k} - \varepsilon_{\bm p} -\omega) \nonumber \\
    && \times (f^0_{\bm k} - f^0_{\bm p}) n_B(\omega) (n_B(\omega)+1) \nonumber \\
    && \times \left[ g(\omega)+f(\omega)\left( J_0(kd)J_0(pd)-J_1(kd)J_1(pd) \right) \right] , \label{spindegtau}
\end{eqnarray}
where $J_n(x)$ is the Bessel function of the first kind of order $n$ and $ {\cal D} = (2 \pi)^{-1}\int  k \partial f_{\bm k}^0 / \partial \varepsilon_{\bm k} dk = -(2 \pi \beta v_F^2)^{-1} \log [ 1 + \exp (\beta \mu) ]$.
%The expression \eqref{spindegtau} defines the collision rate $\tau^{-1}_{\rm tr}$. {}

\begin{figure}[t]
\begin{tabular}{c}
\begin{overpic}[width=0.95\columnwidth]{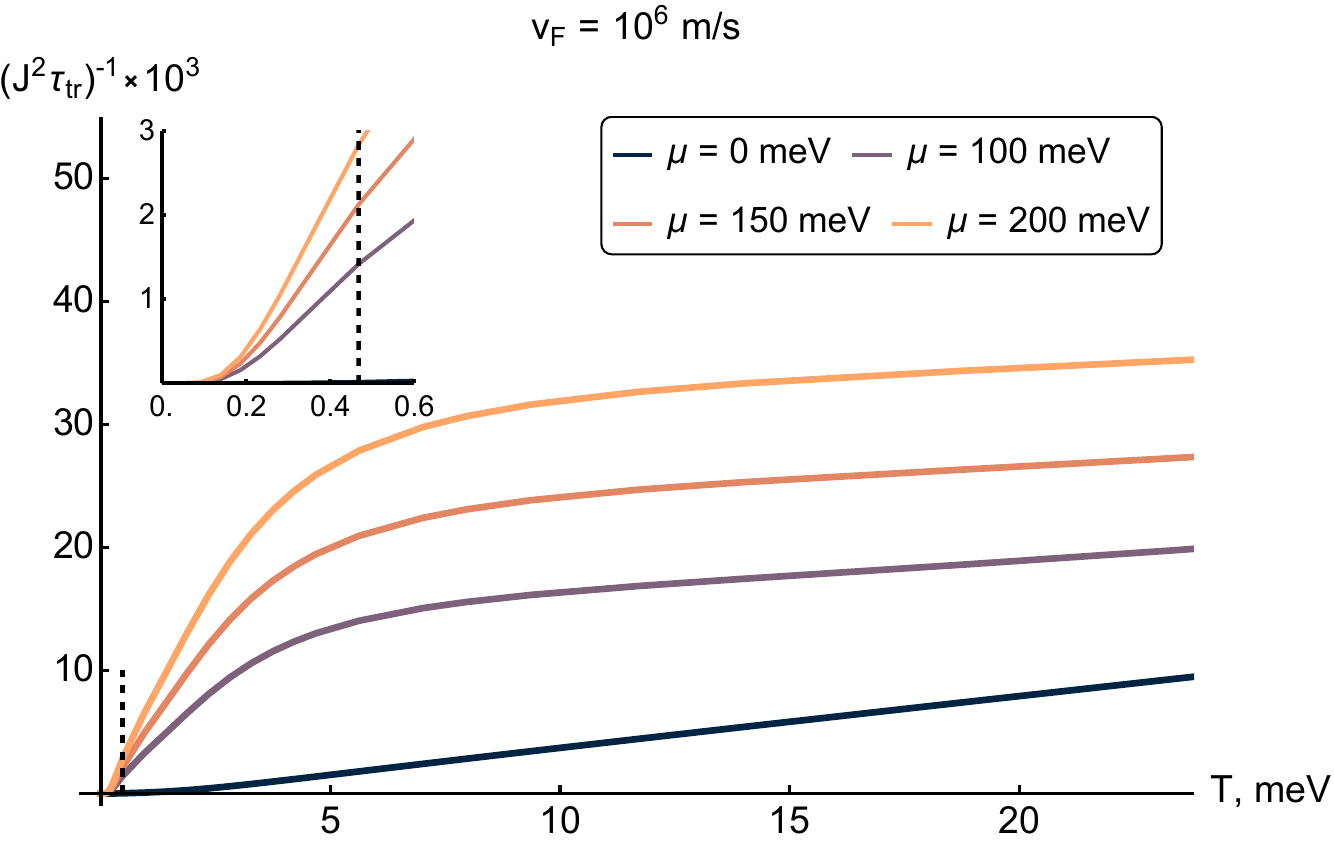}
\put(200,20){{\large (a)}}
\end{overpic}

\\
\begin{overpic}[width=0.95\columnwidth]{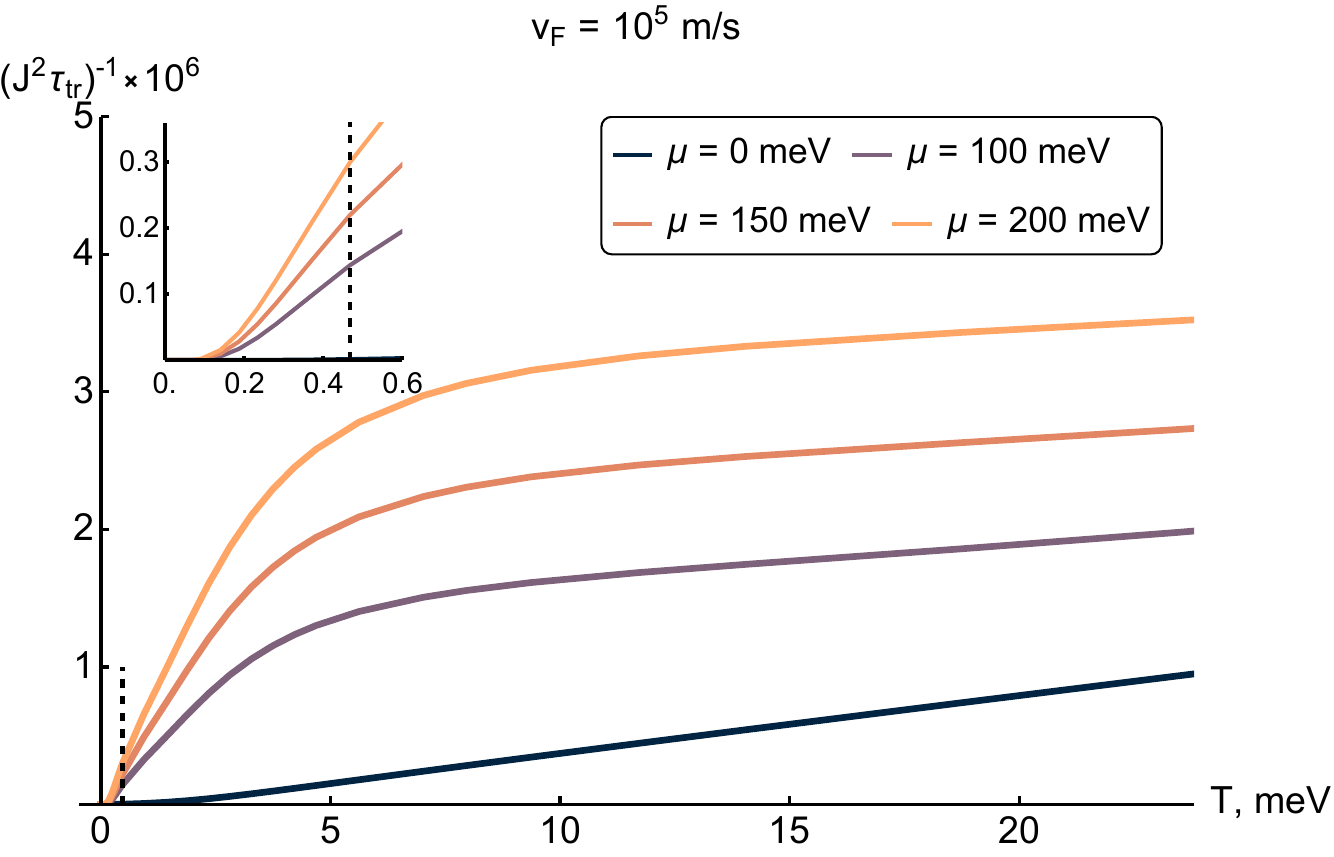}
\put(200,20){{\large (b)}}
\end{overpic}
\end{tabular}
\caption{\label{fig:spindeg300} 
Panel (a): 
The normalized collision rate $(\tau_{\rm tr} J)^{-1}$, defined in Eq.~\eqref{spindegtau}, in the spin-unpolarized case as a function of temperature $T$ for different values of chemical potential $\mu$. A sharp increase of the collision rate is observed at temperature close to the quantum-spin-liquid gap energy. 
Panel (b):
Same as Panel (a), but for a Fermi velocity ten times smaller. The collision rate becomes two orders of magnitudes larger.
{The insets magnifies the region near $k_B T = \Delta$, which is denoted as a dashed vertical line.}
}
\end{figure}

Fig.~\ref{fig:spindeg300}(a) shows the scattering rate $\tau_{\rm tr}$ in units of $J$ as a function of temperature and various chemical potential.  In this plot, the electron Fermi velocity is $v_{\rm F} = 10^6$~m/s as for graphene's massless Dirac fermions. For small chemical potentials, the scattering rate grows linearly with temperature. This is expected since $k_BT$ is the only relevant energy scale in the problem. This universal linear asymptotic behavior is always recovered for high-enough temperatures, when these exceed the chemical potential. 

Increasing the chemical potential, the overall collision rate grows. More importantly, it is nearly zero below the temperature of the order of the quantum-spin-liquid gap.
{This low-temperature activation behavior is arguably the most prominent effect of the QSL on electronic transport, and indicates the gapped nature of flux excitations and hints to the deconfinement of fractionalized excitations~\cite{Knolle2014_PRL_DynamicsKitaev,Banerjee2017_Science_RuCl3_QSL,Do2017_NatPhys_Majorana_RuCl3}. We note that in this range of temperatures the use of the flux-free structure factor
% (calculated in the absence of flux-like excitations) 
is well justified. 
% As shown in Ref.~\cite{XXX}, the gap depends on the magnetic field to the cubic power. Therefore, we expect 
}

{The collision rate} rises sharply {above the flux creation gap} up to temperatures of order of QSL bandwidth, followed by a small dip, to then resume the linear-in-temperature behaviour. {The overall} change in qualitative behaviour, combined with the broad plateaux observed could be used as a signature of frationalized excitations and immobile fluxes in the quantum spin liquid.

In Fig.~\ref{fig:spindeg300}(b) we show the same quantity for lower Fermi velocity $v_{\rm F} = 10^5$ m/s. The scattering rate is two order of magnitude larger, consistent with it scaling with Fermi velocity as $v_F^{-2}$, as expected from inspecting  Eq.~\eqref{spindegtau}. Therefore, materials with a lower Fermi velocity hold higher potential for the experimental observation of quantum-spin-liquid footprints in transport measurements. This also suggests that dilute charges transferred into the $\alpha-$RuCl$_3$ by, e.g., proximity to graphene to dope the very flat d-bands of the material~\cite{Plumb2014_PRB_alphaRuCl3}, {could} experience extremely high collision rates. {Numerical calculations based on realistic band structures of $\alpha-$RuCl$_3$ would be needed to quantify the size of the effect. Such calculations are beyond the scope of the present work.}

\subsection{Spin-unpolarized case II: spin diffusion}
In this section we consider diffusive propagation of spin along the sample. 
{
The experimental setup we have in mind is again similar to that shown in Fig.~\ref{fig:setup}. However, in this case we assume electrodes to be ferromagnetic or heavy-element metals characterized by a strong spin-orbit coupling. In both cases, passing a current in one of the electrodes generates a spin-density accumulation in the metallic channel, which can be in turn detected by the second electrode. 
In the case of ferromagnetic electrodes detection occurs by changing the orientation of magnetization in the injector and detector electrodes in a spin-valve~\cite{Tombros2007_Nature_GrapheneSpinValve,Han2014_NatNano_GrapheneSpintronicsReview} type of configuration.
Passing currents in electrodes characterized by strong spin-orbit coupling results in the generation of a transverse spin current and therefore in the accumulation of spin density in the channel via the spin-Hall effect~\cite{Sinova2015_RMP_SHE,Valenzuela2006_Nature_SHE}. The inverse spin-Hall effect can then be used to detect the spin density reaching a far away electrode by converting it into a transverse voltage drop~\cite{Saitoh2006_APL_ISHE}. 
Hybrid configurations can also be built by combining, e.g. ferromagnetic injectors and inverse-spin-Hall detectors.
To keep the calculation clean, we assume the spin accumulation is generated in the metallic channel, which is therefore separated from the QSL by a small spacer (e.g. an hBN layer). In this way, only electrons contribute to spin diffusion. This allows a clean extraction of the crucial parameter controlling spin diffusion, the spin diffusion time $\tau_{\rm diff}$ (or the spin diffusion length, obtained by multiplying $\tau_{\rm diff}$ by the electron Fermi velocity).
}

{To perform the calculation, we assume that}
at an edge of a sample there is an imbalance between spin up and spin down particles constituting the electric current. This imbalance smears away as the current flows along the sample as the spin-spin interaction tends to randomise the spins. We model this with the following Boltzmann equation 
\be \label{eq:BEspindiff}
    {\bm v}_k \cdot {\bm \nabla}_{\bm r} f_{\bm k \sigma} + e{\bm E} \cdot {\bm \nabla}_{\bm k} f_{\bm k \sigma} = -\frac{\delta f_{\bm k \sigma}}{\tau_{\rm tr}} - \frac{f_{\bm k \sigma}^0 - f_{\bm k \bar{\sigma}}^0}{\tau_{\rm diff}},
\ee
where the full distribution function $f_{\bm k \sigma}$ is approximated as $f_{\bm k \sigma} \approx f_{\bm k \sigma}^0 + \delta f_{\bm k \sigma}$, $\bar{\sigma} = \uparrow$ if ${\sigma} = \downarrow$, and $\bar{\sigma} = \downarrow$ if ${\sigma} = \uparrow$.

The local equilibrium distribution is $ f_{\bm k \sigma}^0 = f^0(\varepsilon_k - \varepsilon^F_\sigma(\bm r))$ where $f^0$ is Fermi-Dirac distribution function and the Fermi energy is $\varepsilon^F_\sigma(\bm r) = \varepsilon^F + \sigma \delta \varepsilon^F (\bm r)$. 
We will be abusing notation by treating $\sigma$ as $\pm 1$ for up and down spins respectively when multiplying it by something else, and as $\sigma = \{ \uparrow, \downarrow \}$ when it serves as an index. 
The diffusion time $\tau_{\rm diff}$ represents the characteristic time scale for relaxation of spin imbalance in the system. 
%and we assume it to be much longer than momentum relaxation $\tau_{\rm tr} \ll \tau_{\rm diff}$. 
We express the perturbation $\delta f_{\bm k \sigma}$ as
\be \label{perturbed}
    \delta f_{\bm k \sigma} = -\tau_{\rm tr} \left( {\bm v}_k \cdot {\bm \nabla}_{\bm r} + e {\bm E} \cdot {\bm v}_k \frac{\partial}{\partial \varepsilon_k} \right) f^0_{\bm k \sigma} .
\ee
We also expand the local equilibrium function as 
\be
    f^0_{\bm k \sigma} = f^0_{\bm k} - \sigma \delta \varepsilon^F (\bm r) \frac{\partial f^0_{\bm k \sigma}}{\partial \varepsilon_k}
\ee
and plug it in \eqref{perturbed}. After defining the spin-dependent electrochemical potential $\mu_\sigma = \varepsilon^F + e \phi + \sigma \delta \varepsilon^F(\bm r)$ we get
\be
    \delta f_{\bm k \sigma} = \tau_{\rm tr} \frac{\partial f^0_{\bm k}}{\partial \varepsilon_k} {\bm v}_{\bm k} \cdot {\bm \nabla}_{\bm r} \mu_\sigma.
\ee
Now let us plug $f_{\bm k \sigma} = f_{\bm k \sigma}^0 + \delta f_{\bm k \sigma}$ into the Boltzmann equation and sum both parts of it over ${\bm k}$ to get
\be \label{eq:spindiff}
    \nabla_{\bm r}^2 \delta \mu = \frac{1}{\nu \tau_{\rm diff}} \delta \mu ,
\ee
where $\delta \mu = \mu_{\uparrow} - \mu_{\downarrow} = 2 \delta \varepsilon^F(\bm r)$ and we have introduced the effective kinematic viscosity $\nu = \tau_{\rm tr} v_F^2/4$. 
%{\color{red}{(the transport time is not necessarily the same entering the viscosity?)}}

To determine the diffusion time $\tau_{\rm diff}$ we again linearise the collision integral and rewrite the kinetic equation to linear order in the perturbations
\be
    \left( \tau_{\text{tr}}(\bm v_{\bm k} \cdot \bm \nabla_{\bm r})^2\mu_{\sigma} + e \bm E \cdot \bm v_{{\bm k}} \right) \frac{\partial f_{\bm k}^0}{\partial \varepsilon_{\bm k}} =  (\partial_t f_{\bm k \sigma})^{\text{lin}}. 
\ee
After summing both sides of the equation above over ${\bm k}$ and $\sigma$ we get 
\begin{widetext}
\be \label{eq:tau_diff}
    \tau_{\text{diff}}^{-1} = \frac{2 J^2}{{\pi \cal D} k_B T} \int \limits_{-\infty}^\infty d \omega \int d k d p \frac{kp}{(2 \pi)^2}
    \delta\left( \varepsilon_{\bm k}-\varepsilon_{\bm p}-\omega \right)
    (g(\omega) + f(\omega) J_0(k d)J_0(p d))
    f^0_{\bm k} (1 - f^0_{\bm p}) (n_B + 1) ,
\ee
\end{widetext}

Equations~\eqref{eq:spindiff} and~\eqref{eq:tau_diff} are derived in App.~\ref{app:diffderivation}.

\begin{figure}[h]
    \includegraphics[width = \linewidth]{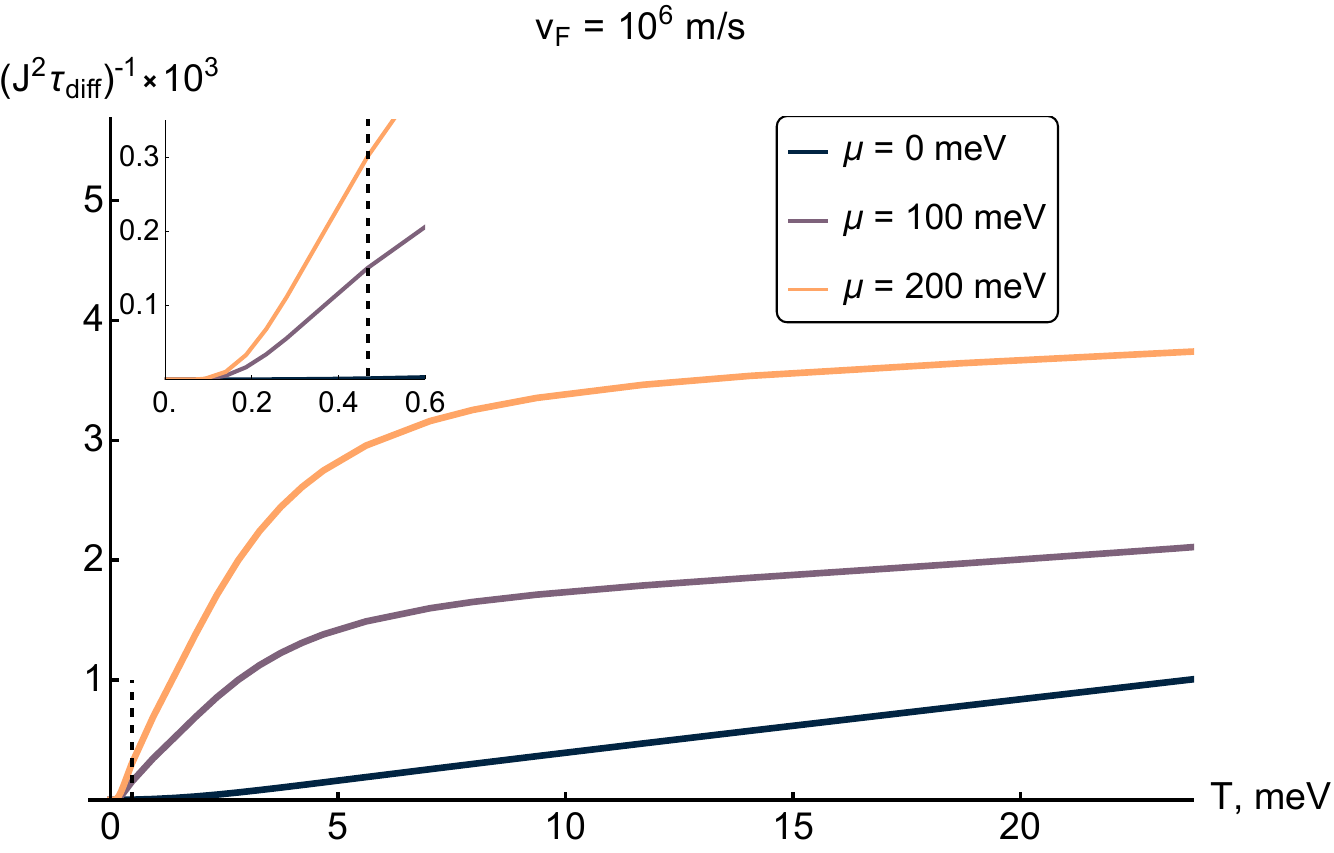} 
    \caption{Diffusion rate Eq.~\eqref{eq:tau_diff} as function of temperature $T$ for different values of chemical potential $\mu$. The black vertical line represents the energy of a flux excitation. {The inset magnifies the region near $k_B T = \Delta$, which is denoted with a dashed vertical line.
    % to guide to the eyes.
    }}
    \label{taudiff}
\end{figure}

Fig.~\ref{taudiff} demonstrates the temperature dependence of diffusion time for different values of $\mu$. One may notice visual similarity between diffusion and transport times, namely $\tau_{\rm tr} \approx 0.1 \tau_{\rm diff} $ for all temperatures. This happens when the QSL bandwidth is much smaller than that of external electrons. In this case the QSL degrees of freedom act like magnetic impurities and the two transport times are now similar. In the limit $T\ll \mu \ll v_F/d$ ($v_F/d\approx W$, where $W$ is graphene's bandwidth) we find $\tau_{\rm tr}/\tau_{\rm diff}=\pi^{-2}$.

\subsection{Spin transport in a spin-polarized conductor}
{One can directly generate a spin currents in the system if bands are (partially) spin-polarized.}
{
The setup we have in mind, with reference always to Fig.~\ref{fig:setup}, is one in which the bands of the metallic channel are spin-split because of, e.g., exchange coupling. 
In this case, graphene would have to be replaced by a different material, for example a magnetic transition metal dichalcogenide~\cite{Bonilla2018_NatNano_VSe2_FM,Nair2019_arXiv_Fe1_3NbS2_switching, He2023_ACSAMI_Graphene_FGT_SpinValve} featuring itinerant electrons exchange-coupled to local magnetic moments.
An intriguing possibility is considering a metallic layer that undergoes a phase transition from spin-unpolarized to spin-split bands. As we show in what follows, This would reveal a sharp transition in the spin-flip time defined below. 
The details of experiments are beyond the scope of the present work, which focuses on the fundamental aspects of the problem.

In this setup, when an electric field is applied, partially spin polarized currents, {\it i.e.} unequal spin-up and -down currents, are produced (see Eq.~(\ref{eq:spin_pol_j_sigma}) below). Unbalanced spin-up and -down currents 
% produce in turn a spin accumulation at the boundary which 
can be detected by ferromagnetic electrodes by tuning the direction of magnetization. 
}

\begin{figure}[t]
\includegraphics[width = 0.75\linewidth]{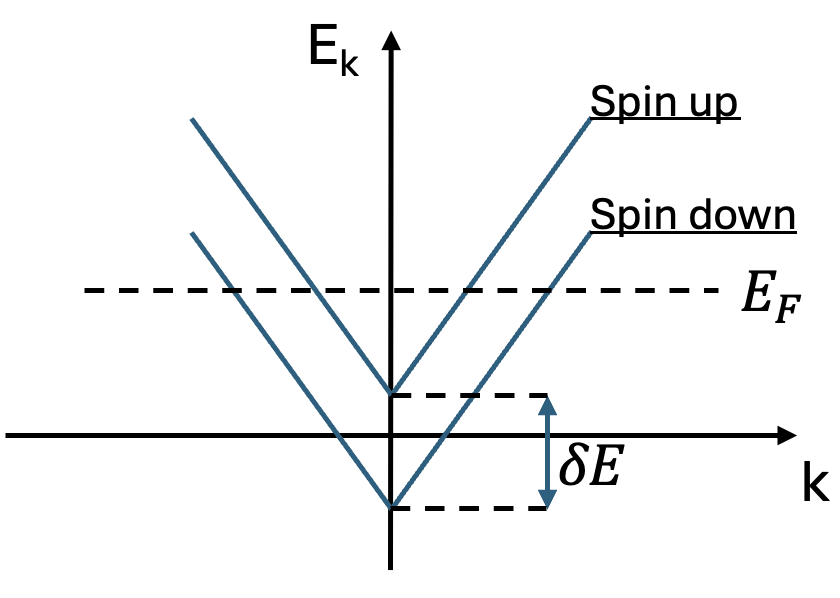}
\caption{\label{fig:dispersion} {Schematic illustration of spin dependent dispersion in a spin polarized conductor: the bands are shifted symmetrically with respect to the zero-field point.} }
\end{figure}

We assume the spin-up and -down bands to be shifted by $\delta\epsilon$ (see Fig.~\ref{fig:dispersion}), {\it i.,e.} $\varepsilon_{\bm k,\uparrow} = \varepsilon_{\bm k,\downarrow}+\delta \varepsilon$. Thus, linearising the energy dispersion around the Fermi level, the spin-up and -down bands, $\varepsilon_{\bm k, \uparrow(\downarrow)} = v_0 |{\bm k }|\pm \delta \varepsilon/2$, are symmetrically shifted with respect to zero energy. The chemical potential $\mu$, measured from the same zero energy level, is the same for both bands. 

In this case, spin-flip processes with the emission of excitations of the quantum spin liquid occur with a typical relaxation time $\tau_{\rm sf}$, which we now determine as a function of temperature and chemical potential. The physics behind $\tau_{\rm sf}$ is quite different from the one governing $\tau_{\rm tr}$. In fact, $\tau_{\rm tr}$ describes the relaxation of deviations in occupation numbers towards the equilibrium state. Thus, it aims at minimizing both fluctuations  $\delta f_{\bm{k} \uparrow}$ and $\delta f_{\bm{k} \downarrow}$ independently. On the contrary, $\tau_{\rm sf}$ tends to equalize deviations from equilibrium for spin-up and -down particles, which are in principle independent. Essentially, a fluctuation in occupation for one species of spins drives deviations in the occupation function for the other species, such that the difference $\delta f_{\bm{k}\downarrow}-\delta f_{\bm{k}\uparrow}$ is minimized.
To incorporate this physics, we will therefore approximate the system of kinetic equations for distribution functions of spin-up and down particles as 
\begin{equation} \label{eq:BEsystem}
    \left\{ 
    \begin{aligned}
        & -e\bm{E}\cdot\bm{v}_{\bm{k},\uparrow} \frac{\partial f^0_{\bm{k},\uparrow}}{\partial \varepsilon_{\bm k, \uparrow}} =-\frac{\delta f_{\bm{k} \uparrow}}{\tau_{\rm tr}}-\frac{\delta f_{\bm{k}\uparrow}-\delta f_{\bm{k}\downarrow}}{\tau_{\rm sf}}, \\
        & -e\bm{E}\cdot\bm{v}_{\bm{k},\downarrow} \frac{\partial f^0_{\bm{k},\downarrow}}{\partial \varepsilon_{\bm k,\downarrow}} =-\frac{\delta f_{\bm{k}\downarrow}}{\tau_{\rm tr}}-\frac{\delta f_{\bm{k}\downarrow}-\delta f_{\bm{k}\uparrow}}{\tau_{\rm sf}}.
    \end{aligned}
    \right.
\end{equation}
Here, $\bm{E}\cdot\bm{v}_{\bm{k},\uparrow(\downarrow)} = E v_0 \cos \phi_{\bm{k}}$, meaning the Fermi velocity is assumed to be the same for both spin-up and -down particles. 
Inverting these equations, we can express perturbations of the distribution functions $\delta f_{\bm{k}\uparrow(\downarrow)}$ in terms of the electric field and of the two relaxation times, $\tau_{\rm tr}$ and $\tau_{\rm sf}$. The resulting expressions and then plugged into Eqs.~\eqref{coll_int_2} and~\eqref{bollzeq}, and a self-consistency condition for the relaxation times is then derived.

To find such condition, we multiply both sides of the two kinetic equations resulting from~\eqref{bollzeq} by $v_{\bm k} \cos \phi_{\bm k}$ and integrate over $\bm k$. It is easy to see that the left-hand side is proportional to $\partial_\mu n_{\sigma} = \partial n_{\sigma}/\partial \mu$.

After some algebra, the kinetic equations can be recast in the following form
\begin{equation}
    \left\{ 
    \begin{aligned}
        &  eEv_0\partial_\mu n_{\uparrow} = A \tau_{\rm tr}+B\tau_{\rm sf},\\
        &  eEv_0\partial_\mu n_{\downarrow} = C \tau_{\rm tr}+D\tau_{\rm sf},
    \end{aligned}
    \right.
\end{equation}
where the coeffcients $A$, $B$, $C$ and $D$ are given in the App.~\ref{app:BIG} alongside the derivation of these formulae. 
Once these coefficients have been determined, the collision rates $\tau_{\rm tr}^{-1}$ and $\tau_{\rm sf}^{-1}$ can be calculated as 
\begin{equation}
    \left\{ 
    \begin{aligned}\label{tausf}
        &  \tau_{\rm tr}^{-1} = \frac{AD-BC}{eEv_0(D \partial_\mu n_{\uparrow} - C \partial_\mu n_{\downarrow})}  ,\\
        &  \tau_{\rm sf}^{-1} = \frac{AD-BC}{eEv_0(A \partial_\mu n_{\downarrow} - B \partial_\mu n_{\uparrow})},
    \end{aligned}
    \right.
\end{equation}
{Using Eq.~(\ref{eq:BEsystem}), spin-up and -down currents are calculated as
\begin{eqnarray} \label{eq:spin_pol_j_sigma}
j^x_\sigma = 
\int \frac{d^2{\bm k}}{(2\pi)^2}
{v}_{{\bm k},\sigma}^x \delta f_{{\bm k}\sigma}
= 
\frac{
e^2 E}
{2}
\frac{
\tau_{\rm sf}}
{1+\kappa^2}
(\partial_\mu n^0_{\bar{\sigma}} + \kappa \partial_\mu n^0_{\sigma}) , \nn \\
%\frac{
%(\tau_{\rm sf} + \tau_{\rm tr}) \partial_\mu n_{\sigma}
%+
%\tau_{\rm tr} \partial_\mu n_{\bar \sigma}}
%{2 + \tau_{\rm sf}/\tau_{\rm tr}}.
%\nonumber\\
\end{eqnarray}
}
{where we have introduced quantity $\kappa = 1+ \tau_{\rm sf}/\tau_{\rm tr}$. }

\begin{figure}[t]
\begin{tabular}{c}
\begin{overpic}[width=0.95\columnwidth]{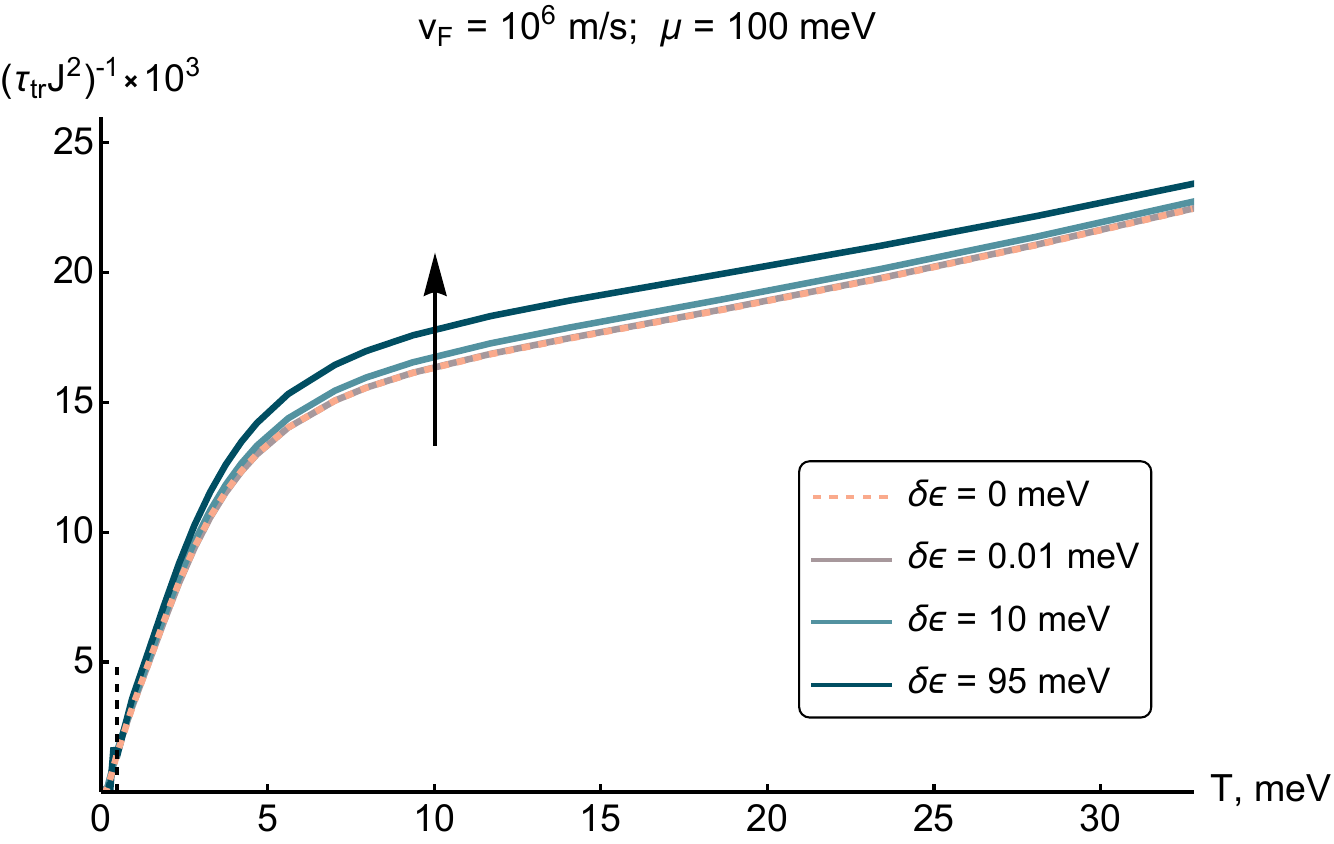}
\put(210,20){{\large (a)}}
\end{overpic}
\\
\begin{overpic}[width=0.95\columnwidth]{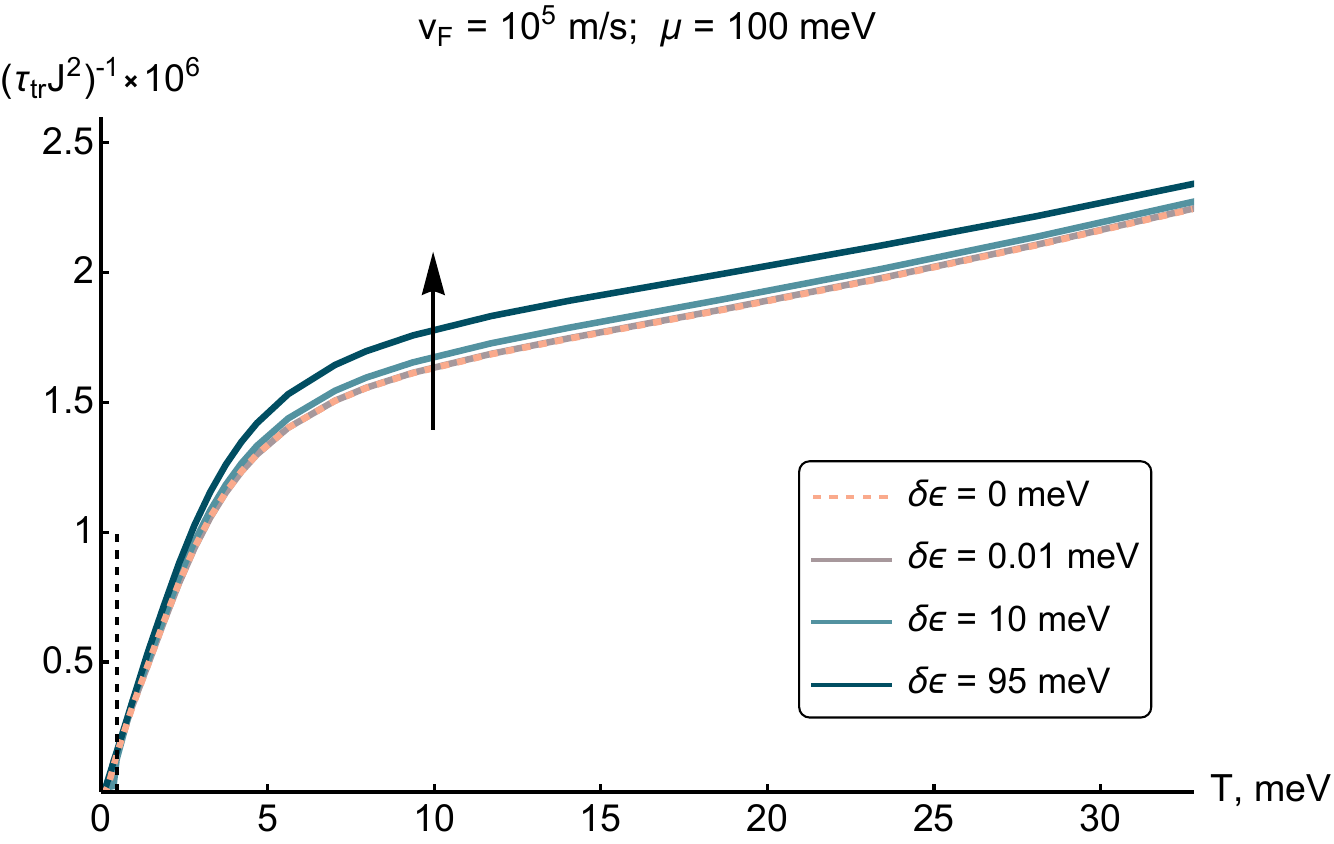}
\put(210,20){{\large (b)}}
\end{overpic}
\end{tabular}
\caption{\label{fig:spinde300}  
Panel (a):
The solid lines are normalized collision rate $(\tau_{\rm tr} J)^{-1}$, Eq.~\eqref{tausf}, for different spin separations $\delta \varepsilon$ as a function of temperature $T$ for a Fermi velocity $v_{\rm F} = 10^6~{\rm m/s}$. The dashed line is the transport time in the spin degenerate case Eq.~\eqref{spindegtau} for the same value of chemical potential $\mu = 100$ meV. 
Panel (b):
Same as Panel (a), but for a reduced Fermi velocity $v_{\rm F} = 10^5~{\rm m/s}$. Arrows indicate direction of curve evolution as we change $\delta \varepsilon$.
}
\end{figure}

The spin polarization has only a negligible effect on the transport relaxation time, as shown in Figs.~\ref{fig:spinde300}(a) and~(b). These show $\tau_{\rm tr}$ as a function of temperature for a fixed chemical potential and various spin polarization (from nearly-unpolarized to almost completely polarized), for the two Fermi velocities used in Figs.~\ref{fig:spindeg300}(a) and~(b), respectively. This insensitivity of the transport time is to be expected, since the transport time relaxes independently fluctuations in the distributions of spin-up and -down electrons towards equilibrium.

\begin{figure}[t]
\includegraphics[width = \linewidth]{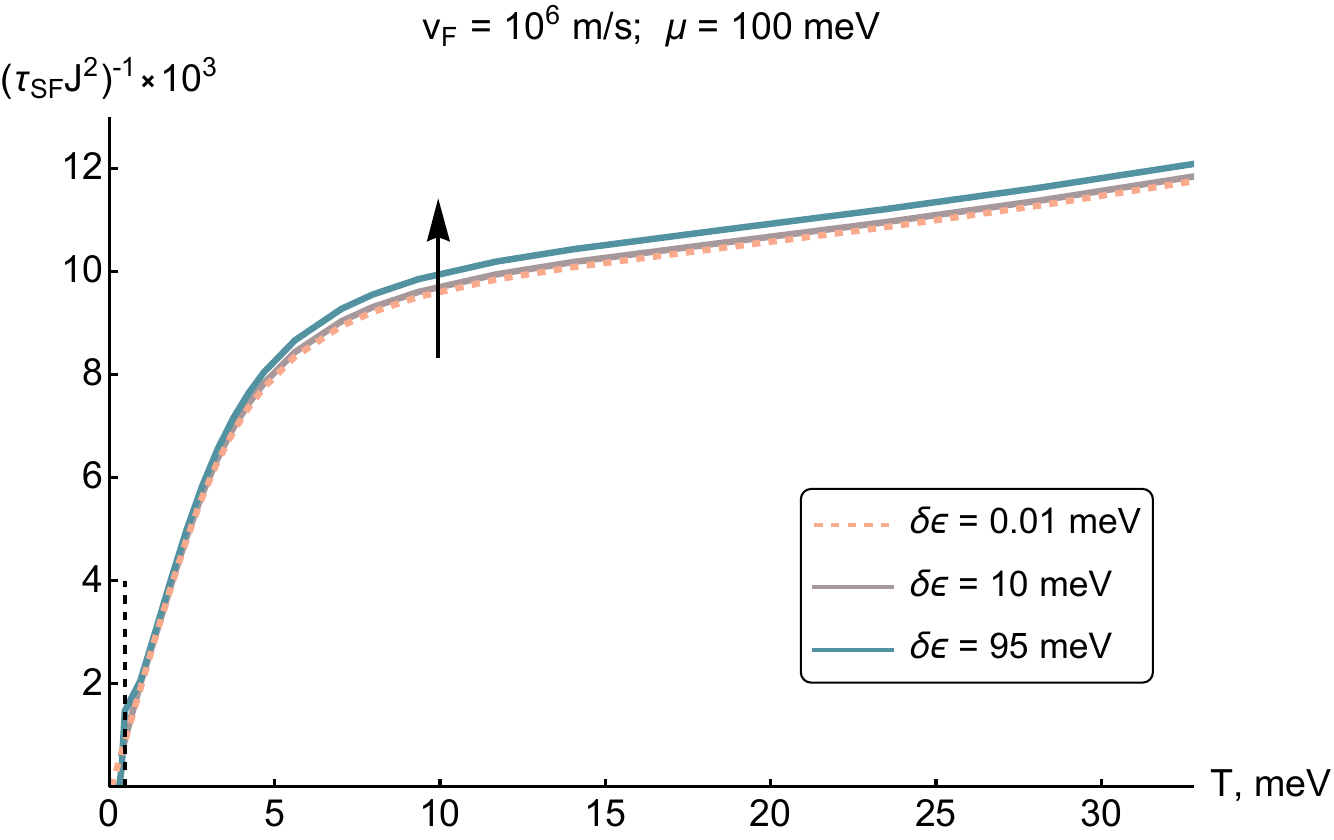}
\caption{\label{fig:spinSFde300} The normalized collision rate $(\tau_{\rm sf} J)^{-1}$, given in Eq.~\eqref{tausf}, for different spin separations $\delta \varepsilon$ as a function of temperature $T$, and for a Fermi velocity $v_{\rm F} = 10^6~{\rm m/s}$. Arrow indicates direction of curve evolution as we change $\delta \varepsilon$. }
\end{figure}

The spin-flip collision rate undergoes a much sharper raise between the completely unpolarized state $\delta\varepsilon = 0$ and any polarized state. In fact, in the completely unpolarized case $\tau_{\rm sf}^{-1} =0$, and it jumps to a finite value, nearly independent of spin polarization, as shown in Fig.~\ref{fig:spinSFde300}, when $\delta\varepsilon$ becomes finite, no matter how small. One should keep in mind that $\tau_{\rm sf}$ does not describe the relaxation of the spin polarization, but of spin currents injected into the system. These are proportional to fluctuations generated {\it on top} of the spin polarization, which depend on the applied electric field but are largely independent of $\delta\varepsilon$. 
However, for $\tau_{\rm sf}$ to become finite it is essential to have a spin polarization, no matter how small, to begin with.

\section{Conclusions and outlook}
In this work, we have developed a kinetic theory to describe charge and spin transport in heterostructures composed of quantum spin liquids (QSLs) and itinerant electron systems, exemplified by $\alpha$-RuCl$_3$ and graphene. By incorporating a Kondo-like interaction between the spin degrees of freedom, we derived a Boltzmann equation and analyzed its collision integral. We studied three different situations. 

Firstly, we considered a completely spin-unpolarized state in which a current is generated by the application of an electric field. We studied how the transport collision rate due to electron-spinon interactions depends on temperature. We found that it is suppressed at low temperatures and it exhibits an activation behavior, consistent with the existence of a gap for spin excitations. Above this threshold, it increases linearly with temperature, and saturate only when this exceeds the bandwith energy (divided by the Boltzmann constant $k_{\rm B}$). The activation threshold and saturation point (which can be experimentally identified by taking the derivative $d\tau_{\rm tr}^{-1}/dT$) can be used to estimate the Kitaev coupling. Furthermore, we found that it strongly depends on the carriers' Fermi velocity, $v_{\rm F}$. When this is reduced, the more frequent interaction between electrons and spins results in a strong increase of the collision rate proportional to $v_{\rm F}^{-2}$.

Secondly, we studied spin diffusion in a spin-unpolarized system where a non-equilibrium spin-imbalanced carrier population is created. We obtained an expression for the spin diffusion rate, which exhibits a similar activation behavior and depends almost linearly with the carriers' chemical potential, consistent with a scaling proportional to the carriers' density of states.

Finally, we consider the interplay between charge and spin transport in a partially spin-polarized conducting system coupled to the Kitaev QSL. We found that the transport time is weakly dependent on spin polarization. Conversely, the spin-flip collision rate jumps from zero to finite, as soon as the system exhibit an infinitesimal spin polarization. However, it remains practically constant for higher degrees of polarization. We also found that, for the chosen parameters, the two collision rates are roughly proportional to each other $\tau_{\rm tr}^{-1}\approx 2 \tau_{\rm SF}^{-1}$.
%$\tau_{\rm tr}/\tau_{\rm diff}=\pi ^{-2}$.

In passing, we note that we corrected the expression for the spin structure factor of the Kitaev model existing in literature~\cite{carrega2020tunneling}.

{The transport rates we calculated in this work can be compared with those obtained from measurements of graphene/$\alpha$-RuCl$_3$ heterostructures (Fig.~\ref{fig:setup}). The transport time $\tau_{\rm tr}$ controls the Drude conductivity $\sigma=ge^2v_0^2 \rho(\mu)\tau_{\rm tr}/2$ (where $g$ is degeneracy factor) or equivalently the longitudinal resistance~\cite{DasSarma2011_RMP_GrapheneTransport,Horng2011_PRB_DrudeWeightGraphene}.
% , measurable via Hall/nonlocal resistance 
The activation behavior of $\tau_{\rm tr}^{-1}$ at $k_{\rm B} T\sim \Delta$ signals the interaction with gapped spin excitations, a consequence of the flux gap. The spin diffusion length 
%$\lambda_s = \sqrt{\nu \tau_{\rm diff}}$ 
can be obtained from the spin-Hall magnetoresistance or from spin-valve effect
\cite{Nakayama2013_PRL_SMR}. Finally, the spin-flip time $\tau_{\rm sf}$ governs spin currents Eq.~\ref{eq:spin_pol_j_sigma}, which 
jump from zero to a finite value in the partially-polarized case. We expect this to be detectable in metallic systems that are either ferromagnets undergoing an ordering phase transition~\cite{Wang2020_NatCommun_FeRhSpinPumping,Weber2021_PRB_Fe1_3NbS2,Deng2018_Nature_FGT_GatedRTFM}, or are exchange-coupled to one. 
{Other mechanisms, for example, magnetic impurities or spin drag due to Coulomb interactions~\cite{PhysRevB.62.4853} could compete with this effect. However, while impurity scattering is expected to be nearly independent of temperature, the electron-spin coupling grows strongly (exponentially) with $T$ around the flux gap $T\approx \Delta/k_{\rm B}$. Furthermore, rates due to scattering against QSL excitations grow with carrier density. On the contrary, the spin drag due to Coulomb interactions is expected to go as~\cite{PhysRevB.91.205423} $(T/\mu)^2$ when $\mu\gg k_{\rm B} T$, and therefore to be suppressed by increasing the carrier density. Therefore, the impact of scattering against QSL excitations could be isolated by studying the resistivity as a function of temperature and electron doping.}
} \\
{
in App.~\ref{app:mag_field} we also 
% report a very small effect of applied magnetic field on transport time $\tau_{\rm tr}$ . For the full range of temperatures, the 
show that a
magnetic field slightly suppresses 
% the 
collision rates. 
% The peak relative difference in collision rates 
The largest effect
is observed at a temperature close to the QSL gap, $T\approx \Delta/k_{\rm B}$. 
% We also observe a 
% delay in 
% shift of t
As the magnetic field is increased,
the
thermal activation of 
% $\tau_{\rm tr}^{-1}$ 
the scattering rate shifts towards higher temperatures.
% in higher magnetic fields. 
This effect is a consequence of the flux gap 
% dependence on magnetic field $H$: $\Delta = \Delta(H)$ 
increasing monotonically with 
% $H$ 
magnetic field~\cite{carrega2020tunneling}. 
} \\
We stress that our results show that the excitations of a Kitaev QSL can be probed in transport experiments performed on graphene/RuCl$_3$ heterostructures. In particular, they can be used to extract parameters of the system, {\it i.e.} as the gap of vison excitations and the bandwidth of matter fermions, both related to the Kitaev coupling. Deviations from the pure QSL behavior due to beyond-Kitaev coupling, whose theoretical description are beyond the scope of this work, can be similarly analyzed.

\begin{acknowledgments}
We are grateful to Ekaterina Nguyen for help with the preparation of Fig.~\ref{fig:setup}. A.E. thanks Alexander Kazantsev for useful discussions. We acknowledge support from the European Commission under the EU Horizon 2020 MSCA-RISE-2019 programme (project 873028 HYDROTRONICS) and from the Leverhulme Trust under the grant agreement RPG-2023-253.
\end{acknowledgments}

%\bibliography{main.bbl}

%\bibliography{refs}
\input{main.bbl}

\begin{widetext}

\appendix

\section{Derivation of the collision integral Eq.~\ref{coll_int}}\label{app:collintder}
We start from the expression for the collision integral 
\be
    \partial_t f_{\bm k\sigma} = \int_0^t dt'\braket{ [V(t),[V(t'),a^{+}_{\bm k\sigma}(t) a_{\bm k\sigma}(t)]] } \label{GoldenRule}.
\ee
In the second order of perturbation theory the time evolution of creation and annihilation operators is given by $a_{{\bm k}\sigma}=e^{-i\varepsilon_{{\bm k} \sigma} t}a_{{\bm k}\sigma}$. Therefore, the inner commutator 
\be
    [V(t'),a^+_{{\bm k}\sigma}a_{{\bm k}\sigma}]=J\sum_{{\bm p}p'\lambda\lambda'}\left[ a^+_{{\bm p}\lambda}e^{i\varepsilon_{{\bm p}\lambda}t'} a_{{\bm p}'\lambda'} e^{-i\varepsilon_{{\bm p}'\lambda'}t'} (\vec{\sigma})_{\lambda\lambda'}\vec{s}^K (\vec{p}-\vec{p}',t'), a^+_{{\bm k}\sigma}a_{{\bm k}\sigma} \right]
\ee
\be
    [a^+_{{\bm p}\lambda}a_{{\bm p}'\lambda'},a^+_{{\bm k}\sigma}a_{{\bm k}\sigma}]=a^+_{{\bm p}\lambda} a_{{\bm p}'\lambda'} (\delta_{\lambda'\sigma} \delta_{{\bm k}p'}-\delta_{\lambda \sigma} \delta_{{\bm k}p})
\ee
\be 
    [V(t'),a^+_{{\bm k}\sigma}a_{{\bm k}\sigma}]=J\sum_{{\bm p}\lambda}\left[ a^+_{{\bm p}\lambda} a_{{\bm k}\sigma} e^{i(\varepsilon_{{\bm p}\lambda}-\varepsilon_{{\bm k}\sigma})t'}(\vec{\sigma})_{\lambda\sigma} \vec{s}^K(\vec{p}-\vec{k,t'}) - \binom{\vec{k}\leftrightarrow \vec{p}}{\lambda \leftrightarrow \sigma}  \right]. 
\ee
%
%{\color{red} ($k$ and $p$ should be boldmath'ed, here and everywhere.)}
{where the symbol $\binom{k\leftrightarrow p}{\lambda \leftrightarrow \sigma}$ stands for the same expression in parenthesis with the symbol, but with all momenta $\vec{p}$ exchanged for $\vec{k}$, all momenta $\vec{k}$ exchanged for $\vec{p}$, all spin indices $\sigma$ exchanged for $\lambda$ and all spin indices $\lambda$ exchanged for $\sigma$. }
%
%{\color{red} stands for... (also why before was only $k\leftrightarrow p$?)}

Then the right-hand side of Eq.~\eqref{GoldenRule} takes the form
\bea \label{eq:goldenrule_1}
    \left( \partial_t f_{{\bm k} \sigma} \right) _{\text{coll}} = 
    J^2\sum_{{\bm p}\lambda}\sum_{{\bm q}{\bm q}'\alpha \alpha'}\int_{-\infty}^t dt' \left[ a^+_{{\bm q}\alpha} a_{{\bm q}'\alpha'} e^{i(\varepsilon_{{\bm q}\alpha}-\varepsilon_{{\bm q}'\alpha'})t} \right. &&(\vec{\sigma})_{\alpha \alpha'}\vec{s}^K(\vec{q}-\vec{q}', t) , a^+_{{\bm p}\lambda} a_{{\bm k}\sigma} e^{i(\varepsilon_{{\bm k}}-\varepsilon_{{\bm k}})t'}(\vec{\sigma})_{\lambda\sigma} \vec{s}^K(\vec{p}-\vec{k,t'}, t')  \nonumber \\
    &&- \left. \binom{\vec{k}\leftrightarrow \vec{p}}{\lambda \leftrightarrow \sigma} \right]
\eea

Rewriting the commutator explicitly, {Eq.~(\ref{eq:goldenrule_1}) becomes}
\begin{align} \label{eq:goldenrule_2}
    \left( \partial_t f_{{\bm k} \sigma} \right) _{\text{coll}} = 
    & J^2\sum_{{\bm p}\lambda}\sum_{{\bm q}\bm q'\alpha \alpha'}\int_{-\infty}^t dt' e^{i(\varepsilon_{{\bm q}\alpha}-\varepsilon_{{\bm q}'\alpha'})t+i(\varepsilon_{{\bm p}\lambda}-\varepsilon_{{\bm k}\sigma})t'} \sigma^\eta_{\alpha \alpha'}\sigma^\gamma_{\lambda \lambda'} \langle a^+_{{\bm q}\alpha}a_{{\bm q}'\alpha'} 
a^+_{{\bm p}\lambda} a_{{\bm k}\sigma} s^\eta_{{\bm q}-\bm q'}(t) s^\gamma_{{\bm p}-\bm k}(t')  \nonumber \\
    &- a^+_{{\bm p}\lambda} a_{{\bm k}\sigma}a^+_{{\bm q}\alpha}a_{{\bm q}'\alpha'} s^\gamma_{{\bm p}-\bm k}(t') s^\eta_{{\bm q}-\bm q'}(t) - \binom{\vec{k}\leftrightarrow \vec{p}}{\lambda \leftrightarrow \sigma} \rangle 
\end{align}
%{\color{red} should there be a square bracket? No it's all fine.}
we then split the expectation values in products of averages for spin operators and creation operators. {Eq.~(\ref{eq:goldenrule_2}) then gives}
\begin{align} \label{eq:goldenrule_3}
    \left( \partial_t f_{{\bm k} \sigma} \right) _{\text{coll}} = 
    & J^2\sum_{{\bm p}\lambda}\sum_{{\bm q}\bm q'\alpha \alpha'}\int_{-\infty}^t dt' e^{i(\varepsilon_{{\bm q}\alpha}-\varepsilon_{{\bm q}'\alpha'})t+i(\varepsilon_{{\bm p}\lambda}-\varepsilon_{{\bm k}\sigma})t'} \sigma^\eta_{\alpha \alpha'}\sigma^\gamma_{\lambda \lambda'} \left[  \langle a^+_{{\bm q}\alpha}a_{{\bm q}'\alpha'} 
a^+_{{\bm p}\lambda} a_{{\bm k}\sigma} \rangle \langle s^\eta_{{\bm q}-\bm q'}(t) s^\gamma_{{\bm p}-\bm k}(t') \rangle \right. \nonumber \\
    & \left. - \langle a^+_{{\bm p}\lambda} a_{{\bm k}\sigma}a^+_{{\bm q}\alpha}a_{{\bm q}'\alpha'} \rangle \langle s^\gamma_{{\bm p}-\bm k}(t') s^\eta_{{\bm q}-\bm q'}(t) \rangle - \binom{\vec{k}\leftrightarrow \vec{p}}{\lambda \leftrightarrow \sigma} \right] 
\end{align}
We us the Wick's theorem for creation operators and obtain
\begin{align}
    & \langle a^+_{{\bm q}\alpha}a_{{\bm q}'\alpha'} \rangle = \delta^{\alpha \alpha'}_{{\bm q} \bm q'}f_{{\bm q}\alpha} \\
    & \langle a_{{\bm q}'\alpha'}a^+_{{\bm q}\alpha} \rangle = \delta^{\alpha \alpha'}_{{\bm q} \bm q'}(1-f_{{\bm q}\alpha})\\
    & \langle a^+_{{\bm q}\alpha}a_{{\bm q}'\alpha'} a^+_{{\bm p}\lambda} a_{{\bm k}\sigma} \rangle = f_{{\bm k}\sigma}\delta^{\alpha \sigma}_{{\bm q} k}(1-f_{{\bm p} \lambda})\delta_{{\bm q}'\bm p}^{\alpha' \lambda} + f_{{\bm q}\alpha}\delta^{\alpha \alpha'}_{{\bm q} \bm q'}f_{{\bm p} \lambda}\delta_{{\bm k} \bm p}^{\sigma \lambda}
\end{align}
We plug this in {Eq.~\eqref{eq:goldenrule_3}}
\begin{align} \label{eq:goldenrule_4}
    \left( \partial_t f_{{\bm k} \sigma} \right) _{\text{coll}} = 
    & J^2\sum_{{\bm p}\lambda}\sum_{{\bm q}q'\alpha \alpha'}\int \limits_{-\infty}^t dt' e^{i(\varepsilon_{{\bm q}\alpha}-\varepsilon_{{\bm q}'\alpha'})t+i(\varepsilon_{{\bm p}\lambda}-\varepsilon_{{\bm k}\sigma})t'} \sigma^\eta_{\alpha \alpha'}\sigma^\gamma_{\lambda \sigma} \left[  \left( f_{{\bm k}\sigma}\delta^{\alpha \sigma}_{{\bm q} \bm k}(1-f_{{\bm p} \lambda})\delta_{{\bm q}'\bm p}^{\alpha' \lambda} + f_{{\bm q}\alpha}\delta^{\alpha \alpha'}_{{\bm q} \bm q'}f_{{\bm p} \lambda}\delta_{{\bm k} \bm p}^{\sigma \lambda} \right) \langle s^\eta_{{\bm q}-\bm q'}(t) s^\gamma_{{\bm p}-\bm k}(t') \rangle \right. \nonumber \\
    & \left. - \left( f_{{\bm p}\lambda}\delta^{\alpha' \lambda}_{{\bm q}' \bm p}(1-f_{{\bm k} \sigma})\delta_{{\bm q}k}^{\alpha \sigma} + f_{{\bm q}\alpha}\delta^{\alpha \alpha'}_{{\bm q} \bm q'}f_{{\bm p} \lambda}\delta_{{\bm k} \bm p}^{\sigma \lambda} \right) \langle s^\gamma_{{\bm p}-\bm k}(t') s^\eta_{{\bm q}-\bm q'}(t) \rangle - \binom{\vec{k}\leftrightarrow \vec{p}}{\lambda \leftrightarrow \sigma} \right]
\end{align}
It is evident that the terms $f_{{\bm q}\alpha}\delta^{\alpha \alpha'}_{{\bm q} \bm q'}f_{{\bm p} \lambda}\delta_{{\bm k} p}^{\sigma \lambda}$ will cancel each other when we take into account {the term} $\binom{\vec{k}\leftrightarrow \vec{p}}{\lambda \leftrightarrow \sigma}$ and use the Kronecker delta's to fix the momenta and spin indices. 
{As shown in Ref.~\cite{carrega2020tunneling}, we can}
% We also
restrict ourselves to the case of diagonal spin-spin correlations $\langle s^\gamma_{{\bm p}-\bm k}(t') s^\eta_{{\bm q}-\bm q'}(t) \rangle=\delta_{\eta \gamma}\langle s^\gamma_{{\bm p}-\bm k}(t') s^\gamma_{{\bm q}-\bm q'}(t) \rangle $. Therefore, {Eq.~(\ref{eq:goldenrule_4})} becomes
\begin{equation} \label{eq:goldenrule_5}
    \left( \partial_t f_{{\bm k} \sigma} \right) _{\text{coll}} = 
    J^2\sum_{{\bm p},\lambda, \gamma} \int \limits_{-\infty}^t dt' e^{i(\varepsilon_{{\bm k}\sigma}-\varepsilon_{{\bm p}\lambda})(t-t')} \sigma^\gamma_{\sigma \lambda}\sigma^\gamma_{\lambda \sigma} \left[ f_{{\bm k}\sigma}(1-f_{{\bm p} \lambda})\langle s^\gamma_{{\bm k}-\bm p}(t) s^\gamma_{{\bm p}-\bm k}(t') \rangle - f_{{\bm p}\lambda}(1-f_{{\bm k} \sigma}) \langle s^\gamma_{{\bm p}-\bm k}(t') s^\gamma_{{\bm k}-\bm p}(t) \rangle \right] - \binom{\vec{k}\leftrightarrow \vec{p}}{\lambda \leftrightarrow \sigma}
\end{equation}
{Since we are interested in the transition rate, we now change the integration variable to} $\xi=t-t'$, send $t \rightarrow\infty$ and use time translation invariance of the correlation functions {to rewrite Eq.~(\ref{eq:goldenrule_5}) explicitly as}
%\begin{equation}
%    J^2\sum_{{\bm p},\lambda, \gamma} \int \limits_{-\infty}^\infty d \xi e^{i(\varepsilon_{{\bm k}\sigma}-\varepsilon_{{\bm p}\lambda})\xi} \sigma^\gamma_{\sigma \lambda}\sigma^\gamma_{\lambda \sigma} \left[ f_{{\bm k}\sigma}(1-f_{{\bm p} \lambda})\langle s^\gamma_{{\bm k}-\bm p}(\xi) s^\gamma_{{\bm p}-\bm k} \rangle - f_{{\bm p}\lambda}(1-f_{{\bm k} \sigma}) \langle s^\gamma_{{\bm p}-\bm k}(-\xi) s^\gamma_{{\bm k}-\bm p} \rangle \right] - \binom{\vec{k}\leftrightarrow \vec{p}}{\lambda \leftrightarrow \sigma}
%\end{equation}
%

\begin{align}
    \left( \partial_t f_{{\bm k} \sigma} \right) _{\text{coll}} = 
    & J^2\sum_{{\bm p},\lambda, \gamma} \int \limits_{-\infty}^\infty d \xi e^{i(\varepsilon_{{\bm k}\sigma}-\varepsilon_{{\bm p}\lambda})\xi} \sigma^\gamma_{\sigma \lambda}\sigma^\gamma_{\lambda \sigma} \left[ f_{{\bm k}\sigma}(1-f_{{\bm p} \lambda})\langle s^\gamma_{{\bm k}-\bm p}(\xi) s^\gamma_{{\bm p}-\bm k} \rangle - f_{{\bm p}\lambda}(1-f_{{\bm k} \sigma}) \langle s^\gamma_{{\bm p}-\bm k}(-\xi) s^\gamma_{{\bm k}-\bm p} \rangle \right] -  \nonumber \\
    & J^2\sum_{{\bm p},\lambda, \gamma} \int \limits_{-\infty}^\infty d \xi e^{-i(\varepsilon_{{\bm k}\sigma}-\varepsilon_{{\bm p}\lambda})\xi} \sigma^\gamma_{\lambda \sigma} \sigma^\gamma_{\sigma \lambda} \left[  f_{{\bm p}\lambda}(1-f_{{\bm k} \sigma}) \langle s^\gamma_{{\bm p}-\bm k}(\xi) s^\gamma_{{\bm k}-\bm p} \rangle - f_{{\bm k}\sigma}(1-f_{{\bm p} \lambda})\langle s^\gamma_{{\bm k}-\bm p}(-\xi) s^\gamma_{{\bm p}-k} \rangle\right]
\end{align}
The top and bottom lines only differ by the sign upon changing $\xi \rightarrow -\xi$ in the second line. {Thus, we can finally rewrite Eq.~(\ref{GoldenRule}) as} 
\begin{equation}
    \left( \partial_t f_{{\bm k} \sigma} \right) _{\text{coll}} = 
    2J^2\sum_{{\bm p},\lambda, \gamma} \int \limits_{-\infty}^\infty d \xi e^{i(\varepsilon_{{\bm k}\sigma}-\varepsilon_{{\bm p}\lambda})\xi} \sigma^\gamma_{\sigma \lambda}\sigma^\gamma_{\lambda \sigma} \left[ f_{{\bm k}\sigma}(1-f_{{\bm p} \lambda})\langle s^\gamma_{{\bm k}-\bm p}(\xi) s^\gamma_{{\bm p}-\bm k} \rangle - f_{{\bm p}\lambda}(1-f_{{\bm k} \sigma}) \langle s^\gamma_{{\bm p}-\bm k}(-\xi) s^\gamma_{{\bm k}-\bm p} \rangle \right]
\end{equation}
Introducing the Fourier transform of the correlation function, 
% (we will distinguish between them by argument): 
\begin{equation*}
    \braket{s^\gamma_{{\bm k}-\bm p}(\xi)s^\gamma_{{\bm p}-\bm k}(0)}=\frac{1}{2\pi}\int \limits_{-\infty}^\infty e^{-i\omega\xi}\braket{s^\gamma_{{\bm k}-\bm p}(\omega)s^\gamma_{{\bm p}-\bm k}}d\omega,
\end{equation*}
We arrive at the following expression for the collision integral Eq.~(\ref{GoldenRule})
\begin{equation}
    \left( \partial_t f_{{\bm k} \sigma} \right) _{\text{coll}} = 
    \frac{J^2}{\pi} \sum_{{\bm p},\lambda, \gamma} \int \limits_{-\infty}^\infty d \omega 
    \delta\left( \varepsilon_{{\bm k}\sigma}-\varepsilon_{{\bm p}\lambda}-\omega \right) 
    \sigma^\gamma_{\sigma \lambda}\sigma^\gamma_{\lambda \sigma} 
    \left[ f_{{\bm k}\sigma}(1-f_{{\bm p} \lambda})\langle s^\gamma_{{\bm k}-\bm p}(\omega) s^\gamma_{{\bm p}-\bm k} \rangle - f_{{\bm p}\lambda}(1-f_{{\bm k} \sigma}) \langle s^\gamma_{{\bm p}-\bm k}(-\omega) s^\gamma_{{\bm k}-\bm p} \rangle \right]
\end{equation}
Next, we evoke the fluctuation-dissipation theorem, {which connects the correlation function $\langle s^\gamma_{{\bm k}-\bm p}(\omega) s^\gamma_{{\bm p}-\bm k} \rangle$ to the imaginary part of the spin-spin response function $\Im \chi^\gamma_{{\bm k}-\bm p} (\omega)$,} to get
\begin{equation}\label{fullCI}
    \left( \partial_t f_{{\bm k} \sigma} \right) _{\text{coll}} = 
    \frac{J^2}{\pi} \sum_{{\bm p},\lambda, \gamma} \int \limits_{-\infty}^\infty d \omega 
    \delta\left( \varepsilon_{{\bm k}\sigma}-\varepsilon_{{\bm p}\lambda}-\omega \right) 
    \sigma^\gamma_{\sigma \lambda}\sigma^\gamma_{\lambda \sigma} 
    \left[ f_{{\bm k}\sigma}(1-f_{{\bm p} \lambda})(n_B(\omega)+1)\Im \chi^\gamma_{{\bm k}-\bm p} (\omega) - f_{{\bm p}\lambda}(1-f_{{\bm k} \sigma})n_B(\omega)\Im \chi^\gamma_{{\bm p}-\bm k}(\omega) \right]
\end{equation}
From Ref.~\cite{carrega2020tunneling}, the response function $\Im \chi^\gamma_{{\bm p}-\bm k}(\omega)$ 
% momentum dependence is just 
depends on momentum according to
$\cos [ (\vec{p}-\vec{k})\cdot\vec{d}^{\gamma} ]$.
% and therefore we get
This allows us to obtain the result from the main text,  Eq.~(\ref{coll_int}).

\section{Linearization of the collision integral in the spin polarized case}
We linearize the collision integral {Eq.~(\ref{coll_int})} by expanding the distribution function as $f_{{\bm k}\sigma} = f_{{\bm k}\sigma}^0+\delta f_{{\bm k}\sigma}$, where $f_{{\bm k}\sigma}^0=[ 1+e^{\beta (\varepsilon_{{\bm k}\sigma}-\mu)} ]^{-1}$ is Fermi distribution. The collision integral \eqref{fullCI} can be rewritten in a more convenient way
\begin{equation} \label{eq:lin_1}
    \left( \partial_t f_{{\bm k} \sigma} \right) _{\text{coll}} = 
    \frac{J^2}{\pi} \sum_{{\bm p},\lambda, \gamma} \int \limits_{-\infty}^\infty d \omega \delta\left( \varepsilon_{{\bm k}\sigma}-\varepsilon_{{\bm p}\lambda}-\omega \right) \sigma^\gamma_{\sigma \lambda}\sigma^\gamma_{\lambda \sigma} \Im \chi^\gamma_{{\bm k}-\bm p} (\omega) (1-f_{{\bm k} \sigma})(1-f_{{\bm p} \lambda})(n_B(\omega)+1) \left[ \frac{f_{{\bm k}\sigma}}{(1-f_{{\bm k} \sigma})} - \frac{f_{{\bm p}\lambda}}{(1-f_{{\bm p} \lambda})}\frac{n_B(\omega)}{n_B(\omega)+1} \right] 
\end{equation}
    % We work out the part with 
    {The product of}
    distribution functions
    {can be simplified to give}
\begin{align}
    & (1-f_{{\bm k} \sigma})(1-f_{{\bm p} \lambda})(n_B(\omega)+1) \left[ \frac{f_{{\bm k}\sigma}}{(1-f_{{\bm k} \sigma})} - \frac{f_{{\bm p}\lambda}}{(1-f_{{\bm p} \lambda})}\frac{n_B(\omega)}{n_B(\omega)+1} \right] = \nonumber \\
    %&(1-f_{{\bm k}\sigma}^0-\delta f_{{\bm k}\sigma})(1-f_{{\bm p} \lambda}^0-\delta f_{{\bm p} \lambda})(n_B(\omega)+1) \left[ \frac{f_{{\bm k}\sigma}^0+\delta f_{{\bm k}\sigma}}{(1-f_{{\bm k} \sigma}-\delta f_{{\bm k}\sigma})} - \frac{f_{{\bm p}\lambda}^0+f_{{\bm p} \lambda}}{(1-f_{{\bm p} \lambda}^0-f_{{\bm p} \lambda})}\frac{n_B(\omega)}{n_B(\omega)+1} \right] = \nonumber \\
    %&\underbrace{(\text{eq.part})}_0 -(n_B(\omega)+1) (\delta f_{{\bm k}\sigma}+\delta f_{{\bm p}\lambda}) \underbrace{ \left[ \frac{f_{{\bm k}\sigma}^0}{(1-f^0_{{\bm k} \sigma})} - \frac{f^0_{{\bm p}\lambda}}{(1-f^0_{{\bm p} \lambda})}\frac{n_B(\omega)}{n_B(\omega)+1} \right] }_{0 \text{ due to energy conservation}} +\nonumber \\
    & (1-f^0_{{\bm k} \sigma})(1-f^0_{{\bm p} \lambda})(n_B(\omega)+1) \left[ \frac{\delta f_{{\bm k}\sigma}}{(1-f_{{\bm k} \sigma}^0)^2} - \frac{\delta f_{{\bm p}\lambda}}{(1-f^0_{{\bm p} \lambda})^2}\frac{n_B(\omega)}{n_B(\omega)} \right] = (TD)_{\sigma \lambda}\left[ \frac{\delta f_{{\bm k}\sigma}}{(1-f_{{\bm k} \sigma}^0)^2} - \frac{\delta f_{{\bm p}\lambda}}{(1-f^0_{{\bm p} \lambda})^2}\frac{n_B(\omega)}{n_B(\omega)+1} \right]
\end{align}
where in the last line we have introduced a shorthand for the prefactor $(TD)_{\sigma \lambda} = (1-f^0_{{\bm k} \sigma})(1-f^0_{{\bm p} \lambda})(n_B(\omega)+1)$ where momentum ${\bm k}$ always comes with spin index $\sigma$ and momentum ${\bm p}$ always comes with index $\lambda$. The linearised collision integral {Eq.~(\ref{eq:lin_1})} is {then rewritten as}
\begin{equation}\label{linCI}
    (\partial f_{{\bm k}\sigma})_{\text{coll}}^{\text{lin}} = \frac{J^2}{\pi} \sum_{{\bm p},\lambda, \gamma} \int \limits_{-\infty}^\infty d \omega \delta\left( \varepsilon_{{\bm k}\sigma}-\varepsilon_{{\bm p}\lambda}-\omega \right) \sigma^\gamma_{\sigma \lambda}\sigma^\gamma_{\lambda \sigma} \Im \chi^\gamma_{{\bm k}-p} (\omega) (TD)_{\sigma \lambda}\left[ \frac{\delta f_{{\bm k}\sigma}}{(1-f_{{\bm k} \sigma}^0)^2} - \frac{\delta f_{{\bm p}\lambda}}{(1-f^0_{{\bm p} \lambda})^2}\frac{n_B(\omega)}{n_B(\omega)+1} \right]
\end{equation}
% We write it down explicitly for spin-up function 
{For $\sigma=\uparrow$, }
after summing over $\gamma$ and $\lambda$ Eq.~(\ref{linCI}) gives 
\begin{align}
  %   & (\partial f_{{\bm k}\uparrow})_{\text{coll}}^{\text{lin}} = \frac{J^2}{\pi} \sum_{{\bm p}} \int \limits_{-\infty}^\infty d \omega \delta\left( \varepsilon_{{\bm k}\uparrow}-\varepsilon_{{\bm p}\downarrow}-\omega \right) \underbrace{\sigma^x_{\uparrow \downarrow}}_1 \underbrace{\sigma^x_{\downarrow \uparrow}}_1 \Im \chi^x_{{\bm k}-p} (\omega) (TD)_{\uparrow \downarrow}\left[ \frac{\delta f_{{\bm k}\uparrow}}{(1-f_{{\bm k} \uparrow}^0)^2} - \frac{\delta f_{{\bm p}\downarrow}}{(1-f^0_{{\bm p} \downarrow})^2}\frac{n_B(\omega)}{n_B(\omega)+1} \right] + \nonumber \\
    % & \frac{J^2}{\pi} \sum_{{\bm p}} \int \limits_{-\infty}^\infty d \omega \delta\left( \varepsilon_{{\bm k}\uparrow}-\varepsilon_{{\bm p}\downarrow}-\omega \right) \underbrace{\sigma^y_{\uparrow \downarrow}}_i \underbrace{\sigma^y_{\downarrow \uparrow}}_{-i} \Im \chi^y_{{\bm k}-p} (\omega) (TD)_{\uparrow \downarrow}\left[ \frac{\delta f_{{\bm k}\uparrow}}{(1-f_{{\bm k} \uparrow}^0)^2} - \frac{\delta f_{{\bm p}\downarrow}}{(1-f^0_{{\bm p} \downarrow})^2}\frac{n_B(\omega)}{n_B(\omega)+1} \right] + \nonumber \\
     %& \frac{J^2}{\pi} \sum_{{\bm p}} \int \limits_{-\infty}^\infty d \omega \delta\left( \varepsilon_{{\bm k}\uparrow}-\varepsilon_{{\bm p}\uparrow}-\omega \right) \underbrace{\sigma^z_{\uparrow \uparrow}}_1 \underbrace{\sigma^z_{\uparrow \uparrow}}_{1} \Im \chi^z_{{\bm k}-p} (\omega) (TD)_{\uparrow \uparrow}\left[ \frac{\delta f_{{\bm k}\uparrow}}{(1-f_{{\bm k} \uparrow}^0)^2} - \frac{\delta f_{{\bm p}\uparrow}}{(1-f^0_{{\bm p} \uparrow})^2}\frac{n_B(\omega)}{n_B(\omega)+1} \right] = \nonumber \\
     (\partial f_{{\bm k}\uparrow})_{\text{coll}}^{\text{lin}} = & \frac{J^2}{\pi} \sum_{{\bm p}} \int \limits_{-\infty}^\infty d \omega \delta\left( \varepsilon_{{\bm k}\uparrow}-\varepsilon_{{\bm p}\downarrow}-\omega \right) \left( \Im \chi^x_{{\bm k}-p} (\omega)+\Im \chi^y_{{\bm k}-p} (\omega) \right) (TD)_{\uparrow \downarrow}\left[ \frac{\delta f_{{\bm k}\uparrow}}{(1-f_{{\bm k} \uparrow}^0)^2} - \frac{\delta f_{{\bm p}\downarrow}}{(1-f^0_{{\bm p} \downarrow})^2}\frac{n_B(\omega)}{n_B(\omega)+1} \right] + \nonumber \\
     & \frac{J^2}{\pi} \sum_{{\bm p}} \int \limits_{-\infty}^\infty d \omega \delta\left( \varepsilon_{{\bm k}\uparrow}-\varepsilon_{{\bm p}\uparrow}-\omega \right) \Im \chi^z_{{\bm k}-p} (\omega) (TD)_{\uparrow \uparrow}\left[ \frac{\delta f_{{\bm k}\uparrow}}{(1-f_{{\bm k} \uparrow}^0)^2} - \frac{\delta f_{{\bm p}\uparrow}}{(1-f^0_{{\bm p} \uparrow})^2}\frac{n_B(\omega)}{n_B(\omega)+1} \right] \label{linspinup}
\end{align}
% And the same for spin-down function
{Conversely, for $\sigma = \downarrow$ Eq.~(\ref{linCI}) gives}
\begin{align}
     & (\partial f_{{\bm k}\downarrow})_{\text{coll}}^{\text{lin}} = \frac{J^2}{\pi} \sum_{{\bm p}} \int \limits_{-\infty}^\infty d \omega \delta\left( \varepsilon_{{\bm k}\downarrow}-\varepsilon_{{\bm p}\uparrow}-\omega \right) \left( \Im \chi^x_{{\bm k}-p} (\omega)+\Im \chi^y_{{\bm k}-p} (\omega) \right) (TD)_{\downarrow \uparrow}\left[ \frac{\delta f_{{\bm k}\downarrow}}{(1-f_{{\bm k} \downarrow}^0)^2} - \frac{\delta f_{{\bm p}\uparrow}}{(1-f^0_{{\bm p} \uparrow})^2}\frac{n_B(\omega)}{n_B(\omega)+1} \right] +  \nonumber \\
     & \frac{J^2}{\pi} \sum_{{\bm p}} \int \limits_{-\infty}^\infty d \omega \delta\left( \varepsilon_{{\bm k}\downarrow}-\varepsilon_{{\bm p}\downarrow}-\omega \right) \Im \chi^z_{{\bm k}-p} (\omega) (TD)_{\downarrow \downarrow}\left[ \frac{\delta f_{{\bm k}\downarrow}}{(1-f_{{\bm k} \downarrow}^0)^2} - \frac{\delta f_{{\bm p}\downarrow}}{(1-f^0_{{\bm p} \downarrow})^2}\frac{n_B(\omega)}{n_B(\omega)+1} \right] \label{linspindown}
\end{align}

\section{Bose summation for the response function Eq.~(\ref{eq:GF1})}\label{app:bosesum}
We start with the response function given in Eq.~(\ref{eq:GF1}) 
\be
    \chi_{ij}^\gamma (i\omega_n) = \int _0^\beta d\tau e^{i\omega_n \tau} \chi_{ij}^\gamma (\tau), \label{fullGF}
\ee
where $\omega_n$ are bosonic Matsubara frequencies. The imaginary-time domain response function is given by \cite{carrega2020tunneling}
\begin{equation} \label{eq:GF1}
    \chi_{\alpha \alpha'}^\gamma (\tau)=[ Q^{1,\gamma}_c(r,\tau,0)-\eta_\alpha \eta_{\alpha'}Q^{1,\gamma}_c(r,0,\tau) ] \delta (r-r'-\delta^\gamma_{\alpha \alpha'}) \braket{TS(\tau, 0)}.
\end{equation}
    
Here we expressed the response function $\chi^\gamma$ in terms of connected Green's functions (GF) $Q_c^{1,\gamma}$ for $f$-fermions Eq.~(\ref{eq:Q_c_gamma_fermions}).
For $\tau>0$, {as in Ref.~\cite{carrega2020tunneling}} we can estimate the $S$-matrix as 
\begin{equation*}
    \braket{TS(\tau, 0)}=e^{-\Delta \tau}, \quad \tau>0.
\end{equation*}

{We note that $Q^{1,\gamma}_c(r,\tau,0)$} 
% the connected GF 
is by definition a GF for fermionic operators. As such, it has to satisfy the anti-periodicity condition. Therefore, we take the Fourier transform %{\color{red} of Eq.~(\ref{xx})} 
using the fermionic Matsubara frequencies $\varepsilon_{n'}$
\be \label{eq:GF2}
    [ Q^{1,\gamma}_c(r,\tau,0)-\eta_\alpha \eta_{\alpha'}Q^{1,\gamma}_c(r,i\varepsilon_{n'}) ]=\beta ^{-1}\sum_{\varepsilon_{n'}} e^{-i\varepsilon_{n'}\tau} [ Q^{1,\gamma}_c(r,\tau,0)-\eta_\alpha \eta_{\alpha'}Q^{1,\gamma}_c(r,-i\varepsilon_{n'}) ] .
\ee
The response function \eqref{fullGF} is bosonic and therefore periodic in imaginary time, hence the $S$-matrix expectation value has to be anti-periodic, meaning 
\begin{equation}
    \braket{TS(\tau -\beta, 0)}=-\braket{TS(\tau, 0)}, \quad \text{for } 0<\tau<\beta.
\end{equation}
We will call $g(\tau)=\braket{TS(\tau, 0)}$ for shorthand. Anti-periodicity again implies that the Fourier transform {of $g(\tau)$ must} contain {only} fermionic (odd) frequencies $\varepsilon_k$
\begin{equation}
    g(\tau)=\frac{1}{\beta}\sum_{\varepsilon_k}e^{-i \varepsilon_k \tau}g_k, \label{gft}
\end{equation}
where $g_k$ are Fourier coefficients which 
% can be fully deduced from 
{are given in terms of the function $g(t)$ in the region}
$\tau>0$ 
{\it i.e.}
% region 
\begin{equation*}
    g_k=\int_0^\beta e^{i \varepsilon_k \tau}g(\tau)d\tau=\frac{1+e^{-\beta \Delta}}{\Delta-i\varepsilon_k}.
\end{equation*}
% Moreover, we can use 
The Fourier representation~\eqref{gft} 
{allows us}
to easily 
% get the right continuation of 
{obtain}
$g(\tau)$ for negative $\tau$:
\begin{align}
    \tau>0: \quad g(\tau)&=\beta^{-1} \sum_{\varepsilon_k=-\infty}^\infty e^{-i\varepsilon_k \tau}\frac{1+e^{-\beta \Delta}}{\Delta-i\varepsilon_k} , \\
    -\tau<0: \quad g(-\tau)&=\beta^{-1} \sum_{\varepsilon_k=-\infty}^\infty e^{+i\varepsilon_k \tau}\frac{1+e^{-\beta \Delta}}{\Delta-i\varepsilon_k}. \label{eq:neg_tau}
\end{align}
We relabel $\varepsilon_k\rightarrow -\varepsilon_k$ in 
%the second line {of 
Eq.~(\ref{eq:neg_tau}) to combine the formulae above in the full $S$-matrix expression
\begin{equation} \label{eq:S_matsubara}
    \braket{TS(\tau, 0)}=\frac{1+e^{-\beta \Delta}}{\beta}\sum_{\varepsilon_k=-\infty}^\infty e^{-i\varepsilon_k \tau} \left( \frac{\theta(\tau)}{\Delta-i\varepsilon_k} + \frac{\theta(-\tau)}{\Delta+i\varepsilon_k} \right)
\end{equation}
{To lighten our notation,} for the time-being we are dropping {the coordinate} $\vec{r}$ in the arguments, {as well as} the overall delta-function $\delta (r-r'-\delta^\gamma_{\alpha \alpha'})$. 
% to write 
{Thus, we rewrite Eq.~(\ref{eq:GF2}) as}
\bea
    [ Q^{1,\gamma}_c(\tau,0)-\eta_\alpha \eta_{\alpha'}Q^{1,\gamma}_c(0,\tau) ] =&& \beta^{-1}\sum_{\epsilon_{n'}}e^{-i \epsilon_{n'} \tau} \left[ \theta(\tau)(Q^{1,\gamma}_c(i \epsilon_{n'})-\eta_\alpha \eta_{\alpha'}Q^{1,\gamma}_c(-i \epsilon_{n'}))+ \right. \nonumber \\
     && +\theta(-\tau)(Q^{1,\gamma}_c(-i \epsilon_{n'})-\eta_\alpha \eta_{\alpha'}Q^{1,\gamma}_c(i \epsilon_{n'})) ].
\eea
The full response function Eq.~(\ref{eq:GF1}), according to Eq.~(\ref{eq:S_matsubara}), is then 
\bea \label{eq:GF3}
    \chi_{\alpha \alpha'}^\gamma (\tau)=\frac{1+e^{-\beta \Delta}}{\beta^2}\sum_{\varepsilon_k, \epsilon_{n'}}&&e^{-i\epsilon_{n'}\tau}e^{-i\varepsilon_{{\bm k}}\tau} \left( \theta(\tau)\frac{Q^{1,\gamma}_c(i \epsilon_{n'})-\eta_\alpha \eta_{\alpha'}Q^{1,\gamma}_c(-i \epsilon_{n'})}{\Delta-i\varepsilon_k}+ \right. \nonumber \\
    && \left. \theta(-\tau)\frac{Q^{1,\gamma}_c(-i \epsilon_{n'})-\eta_\alpha \eta_{\alpha'}Q^{1,\gamma}_c(i \epsilon_{n'})}{\Delta+i\varepsilon_k} \right) .
\eea
Now we take the Fourier transform {of Eq.~(\ref{eq:GF3}) and get}
\bea
    &&\chi_{\alpha \alpha'}^\gamma (i\omega_n)=\frac{1}{2}\int_{-\beta}^\beta d\tau e^{i\omega_n \tau}\chi_{\alpha \alpha'}^\gamma (\tau) = \label{firstline}\\
    &&\frac{1+e^{-\beta \Delta}}{2\beta^2}\sum_{\varepsilon_k, \epsilon_{n'}}\frac{e^{i\beta (\omega_n-\epsilon_{n'}-\varepsilon_k)}}{i(\omega_n-\epsilon_{n'}-\varepsilon_k)}\frac{Q_c^{1,\gamma}(i\epsilon_{n'})-\eta_\alpha \eta_{\alpha'}Q_c^{1,\gamma}(-i\epsilon_{n'})}{\Delta-i\varepsilon_k}+ \nonumber\\
    &&\frac{1+e^{-\beta \Delta}}{2\beta^2}\sum_{\varepsilon_k, \epsilon_{n'}}\int_0^\beta d\tau e^{i(\omega_n-\epsilon_{n'}-\varepsilon_k)\tau}e^{i(\omega_n-\epsilon_{n'}-\varepsilon_k)\beta} \frac{Q^{1,\gamma}_c(-i \epsilon_{n'})-\eta_\alpha \eta_{\alpha'}Q^{1,\gamma}_c(i \epsilon_{n'})}{\Delta+i\varepsilon_k} = \label{secondline} \\
    &&\frac{1+e^{-\beta \Delta}}{2\beta^2}\sum_{\varepsilon_k, \epsilon_{n'}}\beta \delta{(\omega_n-\epsilon_{n'}-\varepsilon_k)}\frac{Q_c^{1,\gamma}(i\epsilon_{n'})-\eta_\alpha \eta_{\alpha'}Q_c^{1,\gamma}(-i\epsilon_{n'})}{\Delta-i\varepsilon_k}+ \nonumber\\
    && \frac{1+e^{-\beta \Delta}}{2\beta^2}\sum_{\varepsilon_k, \epsilon_{n'}} \int_0^\beta d\tau e^{i(\omega_n+\epsilon_{n'}-\varepsilon_k)\tau}\frac{Q_c^{1,\gamma}(i\epsilon_{n'})-\eta_\alpha \eta_{\alpha'}Q_c^{1,\gamma}(-i\epsilon_{n'})}{\Delta+i\varepsilon_k} = \label{thirdline}\\
    &&\frac{1+e^{-\beta \Delta}}{2\beta}\sum_{\epsilon_{n'}} (Q_c^{1,\gamma}(i\epsilon_{n'})-\eta_\alpha \eta_{\alpha'}Q_c^{1,\gamma}(-i\epsilon_{n'}))\left( \frac{1}{\Delta-i\omega_n+i\epsilon_{n'}} + \frac{1}{\Delta+i\omega_n+i\epsilon_{n'}} \right) . \nonumber
\eea
Here, we used the Heaviside theta-functions to split the integral {over $\tau$} into two parts, and shifted $\tau \rightarrow \tau+\beta$ to get from \eqref{firstline} to \eqref{secondline}. Then, we used 
% the definitions for Matsubara frequencies (
that
$e^{i\beta \omega_n}=1$, $e^{i\beta \epsilon_{n'}}=-1$ and $e^{i\beta \varepsilon_k}=-1$) to simplify the expressions. Finally, we have changed the order of summation $\epsilon_{n'}\rightarrow -\epsilon_{n'} $ to get from \eqref{secondline} to \eqref{thirdline}.

We convert the sum over fermionic frequencies {in Eq.~(\ref{thirdline})} into a contour integral {and obtain}
\begin{equation} \label{eq:contour}
    \chi_{\alpha \alpha'}^\gamma (i\omega_n) = \frac{1+e^{-\beta \Delta}}{2}\oint \frac{dz}{2 \pi i} n_F(z)[Q_c^{1,\gamma}(z)-\eta_\alpha \eta_{\alpha'}Q_c^{1,\gamma}(-z)]\left( \frac{1}{\Delta-i\omega_n+z} + \frac{1}{\Delta+i\omega_n+z} \right)
\end{equation}
Where the integration contour is shown {in Fig.~\ref{contour}}.\\

\begin{figure}[t]\centering
    \includegraphics[width=0.95\textwidth]{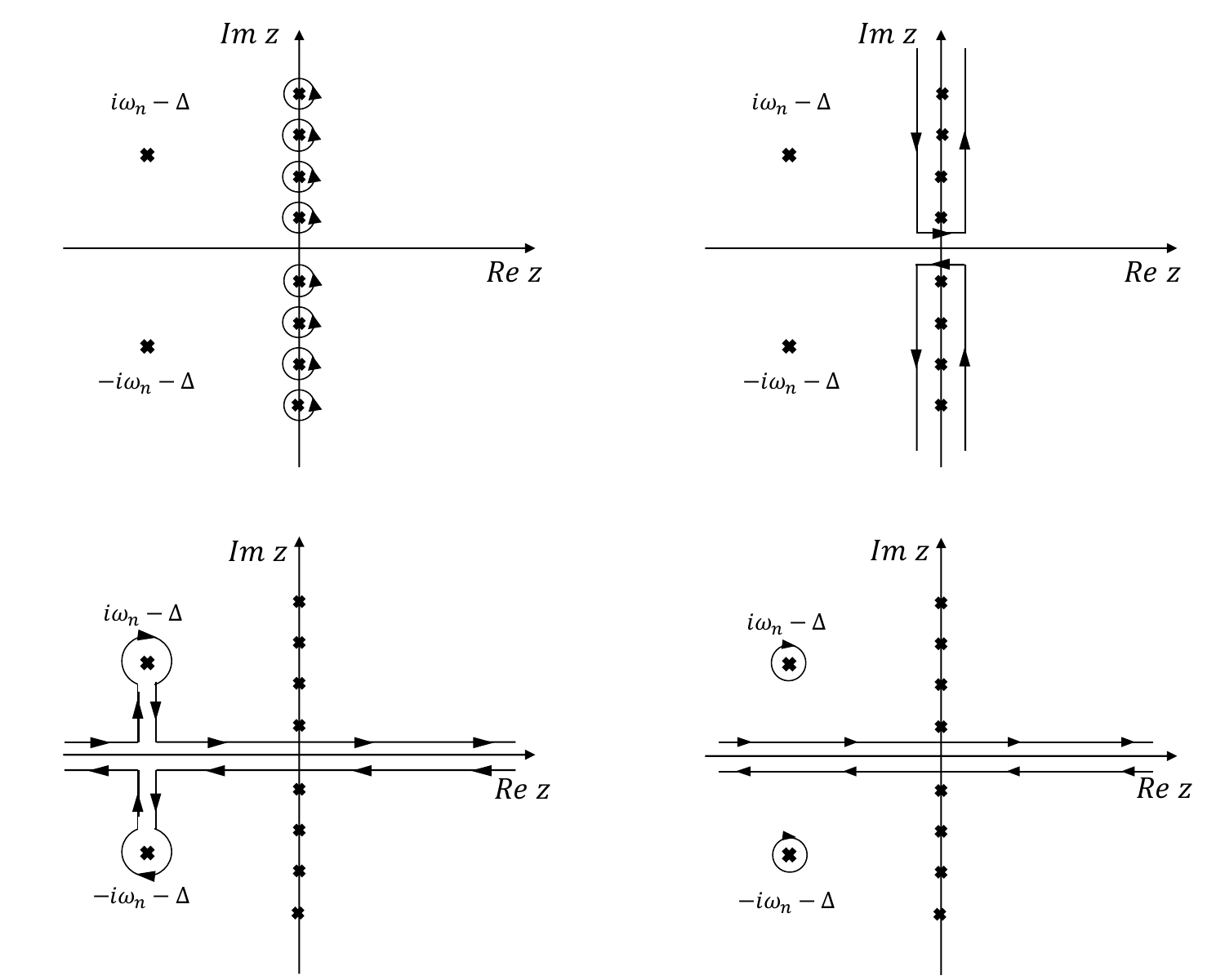}
    \caption{Deformation of the integration contour.}
    \label{contour}
\end{figure}

% According to the deformations of the contour w
We can split the integral {in Eq.~(\ref{eq:contour})} into a sum of two simple poles and two integrals {over branch cuts, {\it i.e.}} along the lines above and below $\mathrm{Re}z=0$:
\bea
    \chi_{\alpha \alpha'}^\gamma (i\omega_n) = &&\frac{1+e^{-\beta \Delta}}{2}\left\{ -n_F(i\omega_n-\Delta) [Q_c^{1,\gamma}(i\omega_n-\Delta)-\eta_\alpha \eta_{\alpha'}Q_c^{1,\gamma}(-i\omega_n+\Delta)] \right. \label{eq:GF4}-\\
    &&  n_F(-i\omega_n-\Delta) [Q_c^{1,\gamma}(-i\omega_n-\Delta)-\eta_\alpha \eta_{\alpha'}Q_c^{1,\gamma}(i\omega_n+\Delta)]   +\nonumber\\
    && \int_{-\infty}^\infty \frac{d\epsilon'}{2 \pi i}n_F(\epsilon'+i0) [Q_c^{1,\gamma}(\epsilon'+i0)-\eta_\alpha \eta_{\alpha'}Q_c^{1,\gamma}(-\epsilon'-i0)]\left( \frac{1}{\Delta-i\omega_n+\epsilon'} + \frac{1}{\Delta+i\omega_n+\epsilon'} \right)- \nonumber \\
    &&  \int_{-\infty}^\infty \frac{d\epsilon'}{2 \pi i}n_F(\epsilon'-i0) [Q_c^{1,\gamma}(\epsilon'-i0)-\eta_\alpha \eta_{\alpha'}Q_c^{1,\gamma}(-\epsilon'+i0)]\left( \frac{1}{\Delta-i\omega_n+\epsilon'} + \frac{1}{\Delta+i\omega_n+\epsilon'} \right)  \nonumber 
\eea
We notice the imaginary parts occurring in the last two lines of Eq.~(\ref{eq:GF4}) can be related by the Schwarz reflection principle $\Im [Q(z)]=-\Im [Q(\Bar{z})]$. From the Eq.~(\ref{eq:GF4})
we get 
\bea \label{eq:GF5}
    \chi_{\alpha \alpha'}^\gamma (i\omega_n) = && \frac{1+e^{-\beta \Delta}}{2}\left\{ -n_F(-\Delta) [Q_c^{1,\gamma}(i\omega_n-\Delta)-\eta_\alpha \eta_{\alpha'}Q_c^{1,\gamma}(-i\omega_n+\Delta) + Q_c^{1,\gamma}(-i\omega_n-\Delta)-\eta_\alpha \eta_{\alpha'}Q_c^{1,\gamma}(i\omega_n+\Delta)]  \right. +\nonumber \\
    && \left. \int_{-\infty}^\infty \frac{d\epsilon'}{\pi}n_F(\epsilon')\Im [Q_c^{1,\gamma}(\epsilon'+i0)+\eta_\alpha \eta_{\alpha'}Q_c^{1,\gamma}(-\epsilon'+i0)] \left( \frac{1}{\Delta-i\omega_n+\epsilon'} + \frac{1}{\Delta+i\omega_n+\epsilon'} \right) \right\}
\eea
We analytically continue the response function {in Eq.~(\ref{eq:GF5})} to real frequencies, $i\omega_n\rightarrow\omega+i\eta$, where $\eta$ is positive infinitesimally small, {and obtain}
\bea
    \chi_{\alpha \alpha'}^\gamma (\omega)=&&\frac{1+e^{-\beta \Delta}}{2}\left\{ -n_F(-\Delta) [Q_c^{1,\gamma}(\omega-\Delta+i\eta)-\eta_\alpha \eta_{\alpha'}Q_c^{1,\gamma}(-\omega+\Delta-i\eta) +   \right.\nonumber \\
    &&Q_c^{1,\gamma}(-\omega-\Delta-i\eta)-\eta_\alpha \eta_{\alpha'}Q_c^{1,\gamma}(\omega+\Delta+i\eta)]+  \\
    && \left. \int_{-\infty}^\infty \frac{d\epsilon'}{\pi}n_F(\epsilon')\Im [Q_c^{1,\gamma}(\epsilon'+i\eta)+\eta_\alpha \eta_{\alpha'}Q_c^{1,\gamma}(-\epsilon'+i\eta)] \left( \frac{1}{\Delta-\omega+\epsilon'-i\eta} + \frac{1}{\Delta+\omega+\epsilon'+i\eta} \right) \right\} \nn
\eea
Next we take the imaginary part {of this expression, and we get}
\bea
    \Im \chi_{\alpha \alpha'}^\gamma (\omega) =&& \frac{1+e^{-\beta \Delta}}{2}\left\{-n_F(-\Delta) \Im [Q_c^{1,\gamma}(\omega-\Delta+i\eta)+\eta_\alpha \eta_{\alpha'}Q_c^{1,\gamma}(-\omega+\Delta+i\eta) +   \right.\nonumber \\
    &&-Q_c^{1,\gamma}(-\omega-\Delta+i\eta)-\eta_\alpha \eta_{\alpha'}Q_c^{1,\gamma}(\omega+\Delta+i\eta)]+ \nn \\
    && \left. \int_{-\infty}^\infty \frac{d\epsilon'}{\pi}n_F(\epsilon')\Im [Q_c^{1,\gamma}(\epsilon')+\eta_\alpha \eta_{\alpha'}Q_c^{1,\gamma}(-\epsilon')] \left( \pi \delta(\Delta-\omega +\epsilon')-\pi \delta(\Delta+\omega +\epsilon') \right) \right\} = \\
    && \frac{1+e^{-\beta \Delta}}{2} \left\{  \Im [Q_c^{1,\gamma}(\omega-\Delta)+\eta_\alpha \eta_{\alpha'}Q_c^{1,\gamma}(-\omega+\Delta) ](n_F(\omega-\Delta)-n_F(-\Delta))- \right. \nn \\
    && \left. \Im [Q_c^{1,\gamma}(-\omega-\Delta)+\eta_\alpha \eta_{\alpha'}Q_c^{1,\gamma}(\omega+\Delta) ](n_F(-\omega-\Delta)-n_F(-\Delta)) \right\} 
\eea
Now we restore all the dropped {spatial dependence} 
% arguments 
and the delta-function to get 
% the final answer
\bea
     \Im \chi_{\alpha \alpha'}^\gamma (\vec{r},\vec{r'},\omega)=&&\frac{1}{2}(1+e^{-\beta \Delta})\delta(\vec{r}-\vec{r'}-\vec{\delta}_{\alpha \alpha'}^\gamma)\times \nn \\ 
    &&\left\{ \Im [Q_c^{1,\gamma}(\vec{r},\omega-\Delta)+\eta_\alpha \eta_{\alpha'}Q_c^{1,\gamma}(\vec{r},-\omega+\Delta) ](n_F(\omega-\Delta)-n_F(-\Delta))- \right. \nn \\
    &&\left. \Im [Q_c^{1,\gamma}(\vec{r},-\omega-\Delta)+\eta_\alpha \eta_{\alpha'}Q_c^{1,\gamma}(\vec{r},\omega+\Delta) ](n_F(-\omega-\Delta)-n_F(-\Delta)) \right\} \label{imq}
\eea
This function is also explicitly anti-symmetric in $\omega$, {\it i.e.} $\Im \chi_{\alpha \alpha'}^\gamma (\vec{r},\vec{r'},-\omega)=-\Im \chi_{\alpha \alpha'}^\gamma (\vec{r},\vec{r'},\omega)$, as expected from an imaginary part of a response function. \\

It can be shown \cite{carrega2020tunneling} that $Q_c^{1,\gamma}(\vec{r},\omega)=Q_c^{1,\gamma}(\omega)$ is independent of the coordinate {${\bm r}$}. We will use this fact in further calculations. 

\subsection{Summation over $\alpha$ and $\alpha'$ {in Eq.~(\ref{imq})}}
% The indices $\alpha, \alpha'={A,B}$ denote the site an operator belongs to.
We sum the response function \eqref{imq} over the indices {$\alpha, \alpha'={A,B}$, which denote the site to which an operator belongs}. Below we explicitly define quantities, which depend on the atom's site within a unit cell
\bea
    &&\vec{\delta}_{\alpha \alpha'}^\gamma = \begin{cases}
        \vec{\delta}_{A A}^\gamma = \vec{\delta}_{B B}^\gamma=0 \\
        \vec{\delta}_{A B}^\gamma = \vec{d}_\gamma \\
        \vec{\delta}_{B A}^\gamma = -\vec{d}_\gamma
    \end{cases}\\
    \vec{d}_x^A=\frac{\sqrt{3} a_{\mathrm{K}}}{3}(-1,0), \quad &&\vec{d}_y^A=\frac{\sqrt{3} a_{\mathrm{K}}}{3}\left(\frac{1}{2},-\frac{\sqrt{3}}{2}\right), \quad \vec{d}_z^A=\frac{\sqrt{3} a_{\mathrm{K}}}{3}\left(\frac{1}{2}, \frac{\sqrt{3}}{2}\right) , \\
    && \eta_A = 1,  \quad \eta_B=-1,
\eea
where $a_K\approx 0.3$ nm \cite{PhysRevB.93.155143} is the lattice spacing. \\
The fact that the only $\vec{r}$ dependence is in the delta-function allows us to find a simple expression for $\Im \chi$ in momentum space. We can take Fourier transform of Eq.~(\ref{imq}) with respect to spatial coordinates and rewrite it in a reduced form featuring the structure of sublattice indices:
\begin{equation} \label{eq:chi_reduced}
    \Im \chi_{\alpha \alpha'}(\omega, \vec{k})=e^{-i \vec{k}\vec{\delta}_{\alpha \alpha'}^\gamma}\{A(X+\eta_\alpha \eta_{\alpha'}Y)+B(Z+\eta_\alpha \eta_{\alpha'}W) \}
\end{equation}
Where $A = \frac{1}{2}(1+e^{-\beta \Delta})(n_F(\omega-\Delta)-n_F(-\Delta))$, $B = \frac{1}{2}(1+e^{-\beta \Delta})(n_F(-\omega-\Delta)-n_F(-\Delta))$, $X = \Im Q_c^{1,\gamma}(\omega - \Delta)$, $Y = \Im Q_c^{1,\gamma}(-\omega + \Delta)$, $Z = \Im Q_c^{1,\gamma}(-\omega - \Delta)$ and $W = \Im Q_c^{1,\gamma}(\omega + \Delta)$. \\
Now we perform the summation {over $\alpha,\alpha'$ in Eq.~(\ref{eq:chi_reduced})}
\bea
    &&\Im \chi(\omega, \vec{k})  =\sum_{\alpha \alpha'}\Im \chi_{\alpha \alpha'}(\omega, \vec{k})=e^{-i \vec{k}\vec{\delta}_{A A}^\gamma} \{ A(X+Y)+B(Z+W) \}+e^{-i \vec{k}\vec{\delta}_{A B}^\gamma} \{ A(X-Y)+B(Z-W) \}+\nn \\
    &&e^{-i \vec{k}\vec{\delta}_{B A}^\gamma} \{ A(X-Y)+B(Z-W) \}+e^{-i \vec{k}\vec{\delta}_{A A}^\gamma} \{ A(X+Y)+B(Z+W) \} =\{ A(X+Y)+B(Z+W) \}+ \nn \\
    &&e^{-i \vec{k}\vec{d}^\gamma} \{ A(X-Y)+B(Z-W) \}+e^{+i \vec{k}\vec{d}^\gamma} \{ A(X-Y)+B(Z-W) \}+ \{ A(X+Y)+B(Z+W) \} =\nn \\
    && (AX + BZ)(2+e^{-i \vec{k}\vec{d}^\gamma}+e^{i \vec{k}\vec{d}^\gamma})+(AY+BW)(2-e^{-i \vec{k}\vec{d}^\gamma}-e^{i \vec{k}\vec{d}^\gamma})= \nn \\
    && 2(AX+BZ)(1+\cos (\vec{k}\vec{d}^\gamma))+2(AY+BW)(1-\cos (\vec{k}\vec{d}^\gamma)).
\eea
Restoring the original functions we get
\bea
    &&\Im \chi(\omega, \vec{k}) = \nn \\
    &&(1+e^{-\beta \Delta})\left\{ [(n_F(\omega-\Delta)-n_F(-\Delta)) \Im Q_c^1(\omega-\Delta) - (n_F(-\omega-\Delta)-n_F(-\Delta)) \Im Q_c^1(-\omega-\Delta)](1+\cos (\vec{k}\vec{d}^\gamma)) +\right. \nn \\
    && \left.  [(n_F(\omega-\Delta)-n_F(-\Delta)) \Im Q_c^1(-\omega+\Delta) - (n_F(-\omega-\Delta)-n_F(-\Delta)) \Im Q_c^1(\omega+\Delta)](1-\cos (\vec{k}\vec{d}^\gamma)) \right\} \label{imchi}
\eea
We note that the imaginary part of the response function is explicitly symmetric in momentum $\Im \chi(\omega, \vec{k})=\Im \chi(\omega, -\vec{k})$.

\section{Spin unpolarised collision rate}\label{app:unpolCR}
The {right-hand side of the} kinetic equation \eqref{eq:unpol_BE_RTA} may be simplified to the following form 
\be
    &&J^2 \int \frac{d^2 k d^2 p d\omega}{(2\pi)^4}\delta(\varepsilon_{\bm k} - \varepsilon_{\bm p} - \omega) (f^0_{\bm k}-f^0_{\bm p}) n_B(\omega) (n_B(\omega)+1) \left[ v_k^x-v_p^x \right]^2\Im \chi(\omega, \vec{k}-\vec{p})= 
    \nn \\
    &&
    \frac{1}{\tau_{\rm tr} \beta }\int \frac{d^2k}{(2\pi)^2}v^2_k\cos^2\phi_{\bm k}\frac{\partial f^0_{\bm k}}{\partial \varepsilon_{\bm k}} \label{simplkineq}
\ee

{We note that} the imaginary part of the response function \eqref{imchi} has 
{a simple angular dependence. It is a sum of two terms, one of which is a constant and the other goes as}
% just two modes with regard to angular variables: $1$ and 
$\cos [ (\vec{k}-\vec{p})\vec{d}^\gamma ] $. The 
% rest of the collision integral has 
{only remaining angular dependence in the collision integral is due to the factor}
$[v_k^x-v_p^x]^2=v_0^2(\cos\phi_k-\cos\phi_p)^2$. {We can }average over the directions $\gamma=\{ x,y,z \}$ {and obtain}

\be
    \braket{\cos [ (\vec{k}-\vec{p})\cdot\vec{d}^\gamma ]}=\frac{1}{3}\sum_\gamma \cos [ (\vec{k}-\vec{p})\cdot\vec{d}^\gamma ]
\ee
% The dot-products will contain 
% \bea
%     && \vec{k}\cdot\vec{d}^z=kd\cos \phi_k ,\nn \\
%     && \vec{k}\cdot\vec{d}^y=kd\cos \left( \phi_k + \frac{2 \pi}{3} \right) , \\
%     && \vec{k}\cdot\vec{d}^x=kd\cos \left( \phi_k - \frac{2 \pi}{3} \right) .\nn 
% \eea    
% So 
We can rewrite the left-hand side of the kinetic equation \eqref{simplkineq} as
\bea
    J^2 v_0^2 \int \frac{d^2 k d^2 p d\omega}{(2\pi)^4}\delta(\varepsilon_k-\varepsilon_{\bm p}-\omega) (f^0_k-f^0_{\bm p}) n_B(\omega) (n_B(\omega)+1)\times \\
     (\cos\phi_k-\cos\phi_p)^2 \left( g(\omega)1+f(\omega)\braket{\cos [ (\vec{k}-\vec{p})\cdot\vec{d}^\gamma ]} \right) 
\eea
where we have introduced 
\bea \label{eq:fandg}
    g(\omega) = &&(1+e^{-\beta \Delta})\left\{ [n_F(\omega - \Delta)-n_F(-\Delta)]\Im[Q_c^1(\omega-\Delta)+Q_c^1(-\omega+\Delta)]- \right. \nn \\
    && \left. [n_F(-\omega - \Delta)-n_F(-\Delta)]\Im[Q_c^1(\omega+\Delta)+Q_c^1(-\omega-\Delta)] \right\}  \\
    f(\omega) = &&(1+e^{-\beta \Delta})\left\{ [n_F(\omega - \Delta)-n_F(-\Delta)]\Im[Q_c^1(\omega-\Delta)-Q_c^1(-\omega+\Delta)]- \right. \nn \\
    && \left. [n_F(-\omega - \Delta)-n_F(-\Delta)]\Im[Q_c^1(-\omega-\Delta)-Q_c^1(\omega+\Delta)] \right\} .
\eea
It can be shown that it's sufficient to evaluate the following {angular} integrals
\bea
    && I_1=\int_0^{2\pi} \frac{d\phi_k d \phi_p}{(2\pi)^4}(\cos\phi_k-\cos\phi_p)^2=\frac{1}{(2\pi)^2}\\
    && I_2(p)=\int_0^{2\pi} \frac{d\phi_k d \phi_p}{(2\pi)^4}\cos \left[ kd\cos\phi_k-pd\cos\phi_p \right] \cos^2\phi_p=\frac{1}{(2\pi)^2}J_0(kd)\left( \frac{J_1(pd)}{pd}-J_2(pd) \right) \\
    && I_2(k)=\int_0^{2\pi} \frac{d\phi_k d \phi_p}{(2\pi)^4}\cos \left[ kd\cos\phi_k-pd\cos\phi_p \right] \cos^2\phi_k=\frac{1}{(2\pi)^2}J_0(pd)\left( \frac{J_1(kd)}{kd}-J_2(kd) \right) \\
    && I_3 = \int_0^{2\pi} \frac{d\phi_k d \phi_p}{(2\pi)^4}\cos \left[ kd\cos\phi_k-pd\cos\phi_p \right](-2\cos\phi_k \cos \phi_p) = -\frac{2}{(2\pi)^2}J_1(kd)J_1(pd)
\eea
% For example:
% \bea
%     && \int_0^{2\pi} \frac{d\phi_k d \phi_p}{(2\pi)^4}\cos \left[ kd\cos (\phi_k-2\pi/3)-pd\cos (\phi_p-2\pi/3) \right] \cos^2\phi_k= \nn\\
%     &&\int_0^{2\pi} \frac{d\phi_k d \phi_p}{(2\pi)^4}\cos \left[ kd\cos \phi_k-pd\cos \phi_p \right] \cos^2(\phi_k+2\pi/3)
% \eea
% which can be then reduced to $I_2(k)$. The result for the example above is
% \begin{equation}
%     \int_0^{2\pi} \frac{d\phi_k d \phi_p}{(2\pi)^4}\cos \left[ kd\cos (\phi_k-2\pi/3)-pd\cos (\phi_p-2\pi/3) \right] \cos^2\phi_k= \frac{\pi}{2 (2\pi)^3}J_0(pd)\left( \frac{4J_1(pd)}{pd} -J_2(pd) \right) .
% \end{equation}
The net result is 
\begin{equation}
    \braket{(\cos \phi_k -\cos \phi_p )^2\cos [ (\vec{k}-\vec{p})\cdot\vec{d}^\gamma ]}=\frac{1}{4\pi^2}\left( J_0(kd)J_0(pd)-J_1(kd)J_1(pd) \right)
\end{equation}
The left hand side {of Eq.~(\ref{simplkineq})} is then 
\begin{equation}
    \frac{J^2 v_0^2}{4\pi^2} \int d k d p d\omega k p \delta(\varepsilon_{\bm k} - \varepsilon_{\bm p} - \omega) (f^0_{\bm k}-f^0_{\bm p}) n_B(\omega) (n_B(\omega)+1)\left[ g(\omega)+f(\omega)\left( J_0(kd)J_0(pd)-J_1(kd)J_1(pd) \right) \right] \label{LHS}
\end{equation}
{Finally,} the kinetic equation ~(\ref{simplkineq}) is then 
\bea \label{RHS}
    && -\frac{\pi}{\tau_{\rm tr} (J v_0 \beta)^2 } \log \left[ 1+e^{\beta \mu} \right] = \label{collrate} \\
    &&\int d k d p d\omega k p \delta(\varepsilon_{\bm k} - \varepsilon_{\bm p} - \omega) (f^0_{\bm k}-f^0_{\bm p}) n_B(\omega) (n_B(\omega)+1)\left[ g(\omega)+f(\omega)\left( J_0(kd)J_0(pd)-J_1(kd)J_1(pd) \right) \right] \nn
\eea

\section{High chemical potential ($\beta\mu \gg 1$) approximation}
First, we start with looking for approximations for the left-hand side (LHS), Eq.~\eqref{LHS}, and right-hand side (RHS), Eq.~\eqref{RHS}, of the kinetic equation under the conditions $\mu \gg \beta^{-1}$ and $\mu \gg \omega_0$. Here, $\mu$ is the electron chemical potential,
% of the electrons, 
$\beta^{-1}=k_BT$ and $\omega_0$ {is the maximum energy of itinerant Majorana fermions, } defined as $\Im \chi (\omega>\omega_0)=0$. \\
Eq.~\eqref{RHS} then gives
% the RHS is 
\begin{equation}
    \frac{1}{4 \pi \beta^2 \tau }\log [1+e^{\beta \mu}] \xrightarrow{\beta \mu\gg 1} \frac{\mu}{4 \pi \tau \beta} . \label{approxRHS}
\end{equation}
In the LHS, Eq.~\eqref{LHS}, we approximate the Fermi distribution functions with Heaviside step-functions
\begin{equation}
    f^0_{\bm k}-f^0_{\bm p} \xrightarrow{\beta \mu\gg 1} \theta(\mu-\varepsilon_{\bm k})-\theta(\mu-\varepsilon_{\bm p}) .
\end{equation}
After using the delta-function to integrate over $\vec{p}$,
% we get
{the right-hand side of Eq.~(\ref{RHS}) gives}
\bea \label{eq:rhs2}
    && \frac{J^2v_0^2}{v_0}\int_0^{\infty}dk \int_{-\omega_0}^{\omega_0}d\omega [ \theta(\mu-v_0 k)-\theta(\mu+\omega -v_0 k) ] k\left( k-\frac{\omega}{v_0} \right) n_B(\omega)(n_B(\omega)+1)F(k,k-\frac{\omega}{v_0},\omega) =  \\
    && J^2 \int_0^{\infty}d\varepsilon \int_{-\omega_0}^{\omega_0}d\omega [ \theta(\mu-\varepsilon)-\theta(\mu+\omega -\varepsilon) ] \frac{\varepsilon}{v_0}\left(\frac{\varepsilon-\omega}{v_0} \right) n_B(\omega)(n_B(\omega)+1)F\left( \frac{\varepsilon}{v_0},\frac{\varepsilon-\omega}{v_0},\omega \right) \nn
\eea
where we changed integration variable to $\varepsilon = v_0 k$. 
The theta-functions {in Eq.~(\ref{eq:rhs2})} cut the integration interval to $\varepsilon_{\bm k}\in [\mu+\omega,\mu]$. Since $\mu \gg \omega $ we will approximate the integral {in Eq.~(\ref{eq:rhs2})} with its mean value 
\begin{equation}
    \int_{\mu+\omega}^{\mu}d\varepsilon_{\bm k} \varepsilon_{\bm k}^2 F \left( \frac{\varepsilon_{\bm k}}{v_0},\frac{\varepsilon_k}{v_0},\omega \right) \approx \mu^2 F(\mu,\mu,\omega)(-\omega) . 
\end{equation}
Overall, the kinetic equation, {Eq.~(\ref{RHS})} now is 
\begin{equation}
    \frac{\mu}{4 \pi \tau \beta} = -\frac{J^2 \mu^2}{v_0^2}\int_{-\omega_0}^{\omega_0}d\omega \omega n_B(\omega)(n_B(\omega)+1) F(\mu,\mu,\omega),
\end{equation}
where 
\begin{equation}
    F(\mu, \mu,\omega) =  g(\omega)+f(\omega) \left[ J_0(\mu d/v_0)^2-J_1(\mu d/v_0)^2 \right] 
\end{equation}
It is easy to check that the function $F(\mu,\mu,\omega)=-F(\mu,\mu,-\omega)$ is antisymmetric in $\omega$. Therefore
\begin{equation}
    -\frac{J^2 \mu^2}{v_0^2}\int_{-\omega_0}^{\omega_0}d\omega 
    % \underbrace{
    \omega
    % }_{\text{asym}} 
    % \underbrace{
    n_B(\omega)(n_B(\omega)+1)
    % }_{sym} \underbrace{
    F(\mu,\mu,\omega)
    % }_{asym}
    =-2\frac{J^2\mu^2}{v_0^2}\int_{0}^{\omega_0}d\omega \omega n_B(\omega)(n_B(\omega)+1) F(\mu,\mu,\omega).
\end{equation}
Another observation is that $d=a_K/\sqrt{3}$ and the corresponding energy scale for this problem is $\Lambda =  v_0/d\approx 2~{\rm eV}$ which is of the order of the linear part of graphene's bandwidth. This means that at low energies $\mu \lesssim 300~{\rm meV}$ the combination 
\begin{equation*}
    J_0(\mu d/v_0)^2-J_1(\mu d/v_0)^2 \approx 1 
\end{equation*}
is just unity and will be omitted in further calculations. Thus, we get for the kinetic equation Eq.~(\ref{RHS})
\begin{equation}
    \frac{\mu}{4 \pi \tau \beta}=-\frac{2 J^2 \mu^2}{(2\pi v_0)^2}\int_{0}^{\omega_0}d\omega \omega n_B(\omega)(n_B(\omega)+1)(g(\omega)+f(\omega)) .
\end{equation}
The collision rate in this approximation is 
\begin{equation}
    \frac{1}{\tau J^2}=-\frac{2}{\pi v_0^2}\frac{\mu}{T}\int_{0}^{\omega_0}d\omega \omega n_B(\omega)(n_B(\omega)+1)(g(\omega)+f(\omega)) \label{approx}
\end{equation}
% The functions $f(\omega)$ and $g(\omega)$ contain convoluted expressions involving $\Im Q_c^1(\pm \omega \pm \Delta)$ and therefore are very hard to be approximated meaningfully. 

\section{Derivation of spin diffusion}\label{app:diffderivation}
We begin with plugging the perturbed distribution function $f_{\bm k \sigma} = f_{\bm k \sigma}^0 + \delta f_{\bm k \sigma}$ into the Boltzmann equation \eqref{eq:BEspindiff}
\be
    \tau_{\rm tr} \frac{\partial f^0_{\bm k}}{\partial \varepsilon_{\bm k}} ({\bm v}_k \cdot{\bm \nabla})^2 \mu_\sigma - \frac{\partial f^0_{\bm k}}{\partial \varepsilon_{\bm k}} ({\bm v}_k \cdot {\bm \nabla}) \mu_\sigma  = - \frac{\partial f^0_{\bm k}}{\partial \varepsilon_{\bm k}} ({\bm v}_k \cdot {\bm \nabla}) \mu_\sigma + \frac{2 \sigma \delta \varepsilon^F }{\tau_{\rm diff}} \frac{\partial f^0_{\bm k}}{\partial \varepsilon_{\bm k}}.
\ee
Summation of both sides over $\bm k$ and $\sigma$ gives equation \eqref{eq:spindiff}. From the latter, to determine the diffusion time $\tau_{\rm diff}$ we again linearize the collision integral given by
\begin{equation}
    (\partial_t f_{\bm k \sigma})_{\rm coll} = \frac{J^2}{\pi} \sum_{p,\lambda, \gamma} \int \limits_{-\infty}^\infty d \omega \delta\left( \varepsilon_{\bm k \sigma}-\varepsilon_{\bm p \lambda}-\omega \right) \sigma^\gamma_{\sigma \lambda}\sigma^\gamma_{\lambda \sigma} \Im \chi^\gamma_{{\bm k- \bm p}} (\omega) \left[ f_{\bm k \sigma}(1-f_{\bm p \lambda})(n_B(\omega)+1) - f_{\bm p \lambda}(1-f_{\bm k \sigma})n_B(\omega) \right] .
\end{equation}
We make the following 
% sequence of
transformation
\be \label{eq:diff_coll}
    f_{k\sigma}(1-f_{p \lambda})(n_B(\omega)+1) - f_{p\lambda}(1-f_{k \sigma})n_B(\omega) = [1 - \delta_{\sigma \lambda}] \frac{\delta \mu}{k_B T} f^0_{\bm k} (1 - f^0_{\bm p}) (n_B + 1)
\ee
%where the last line follows from energy conservation. 
% where
{to obtain which}
we used energy conservation. It {is clear that the term in square brackets on the right-hand side of Eq.~(\ref{eq:diff_coll})} gives zero 
% for the term in square braces
when $\sigma = \lambda$ and 2 when $\sigma \neq \lambda$. {In Eq.~(\ref{eq:diff_coll}), $\delta \mu$ satisfies the diffusion equation Eq.~(\ref{eq:spindiff}) {\it i.e.}}  $\delta \mu = \nu \tau_{\text{diff}} \nabla^2 \delta \mu$. The kinetic equation up to linear terms in perturbations is then
\begin{equation}\label{eq:spindiff2}
    \left( \tau_{\text{tr}}(\bm v_{\bm k} \cdot \bm \nabla)^2\mu_{\sigma}  \right) \frac{\partial f_{\bm k}^0}{\partial \varepsilon_{\bm k}} = (\partial_t f_{\bm k \sigma})^{\text{lin}}_{\rm coll} = 
    \frac{J^2}{\pi} \sum_{p,\lambda, \gamma} \int \limits_{-\infty}^\infty d \omega \delta\left( \varepsilon_{\bm k \sigma}-\varepsilon_{\bm p \lambda}-\omega \right) 
    \sigma^\gamma_{\sigma \lambda} \sigma^\gamma_{\lambda \sigma} 
    \Im \chi^\gamma_{{\bm k- \bm p}} (\omega) 
    (1 - \delta_{\sigma \lambda}) \frac{\delta \mu}{k_B T} f^0_{\bm k} (1 - f^0_{\bm p}) (n_B + 1) .
\end{equation}
We sum Eq.~(\ref{eq:spindiff2}) over $\bm k$ and get 
\bea
    &&\frac{v_F^2 \tau_{\text{tr}}}{2} (\nabla^2 \mu_\sigma)\int dk \frac{k}{2 \pi} \frac{\partial f_{\bm k}^0}{\partial \varepsilon_{\bm k}} = \nn \\
    &&\frac{\sigma \delta \mu J^2}{\pi k_B T}\int \limits_{-\infty}^\infty d \omega \int \frac{d^2 k d^2 p}{(2 \pi)^4} 
    \delta\left( \varepsilon_{\bm k}-\varepsilon_{\bm p}-\omega \right)
    (\Im \chi^x_{{\bm k- \bm p}} (\omega) + \Im \chi^y_{{\bm k- \bm p}} (\omega) )
     f^0_{\bm k}(1 - f^0_{\bm p}) (n_B + 1) .
\eea
We will call $ {\cal D} = \int dk \frac{k}{2 \pi} \frac{\partial f_{\bm k}^0}{\partial \varepsilon_{\bm k}} $. Now we subtract from the equation for spin-up particles ($\sigma = \uparrow$) the equation for spin-down particles ($\sigma = \downarrow$) {and obtain}
\be \label{eq:812}
    \nu {\cal D} \nabla^2 \delta \mu = \delta \mu \frac{J^2}{\pi k_B T} \int \limits_{-\infty}^\infty d \omega \int \frac{d^2 k d^2 p}{(2 \pi)^4} 
    \delta\left( \varepsilon_{\bm k}-\varepsilon_{\bm p}-\omega \right)
    (\Im \chi^x_{{\bm k- \bm p}} (\omega) + \Im \chi^y_{{\bm k- \bm p}} (\omega) )
     f^0_{\bm k} (1 - f^0_{\bm p}) (n_B + 1) .
\ee
We then utilize the representation for the imaginary part of the response function {given in Eq.~(\ref{eq:fandg}), {\it i.e.}} 
\be
    \Im \chi^\gamma_{{\bm k- \bm p}} (\omega) = g(\omega) + f(\omega) \cos [ (\bm k - \bm p) \cdot \bm d^\gamma ] ,
\ee
to perform angular integration in the RHS of \eqref{eq:812}, {whcih gives}
\bea
    \int \limits_{0}^{2 \pi}\frac{d \phi_{\bm k} d \phi_{\bm p}}{(2\pi)^2}\Im \chi^\gamma_{{\bm k- \bm p}} (\omega) = g(\omega) + f(\omega) J_0(k d)J_0(p d),
\eea
where we assumed $\gamma$ to be either $x$ or $y$. The angular average of the response function is the same in these two directions. 
{Finally, we} substitute $\delta \mu = \nu \tau_{\text{diff}} \nabla^2 \delta \mu$ into {Eq.~(\ref{eq:812})} to get the final expression for diffusive time
\be
    \tau_{\text{diff}}^{-1} = \frac{2 J^2}{\pi {\cal D} k_B T} \int \limits_{-\infty}^\infty d \omega \int d k d p \frac{kp}{(2 \pi)^2}
    \delta\left( \varepsilon_{\bm k}-\varepsilon_{\bm p}-\omega \right)
    (g(\omega) + f(\omega) J_0(k d)J_0(p d))
    f^0_{\bm k} (1 - f^0_{\bm p}) (n_B + 1) .
\ee

\section{System of coupled kinetic equations} \label{app:BIG}
In 
% the main text 
Eq.~\ref{eq:BEsystem} we chose the relaxation time approximation to be represented by two time scales, the transport relaxation time $\tau_{\rm tr}$ and spin-flip one $\tau_{\rm sf}$. {This gives the set of equations}
\begin{equation}
    \left\{ 
    \begin{aligned} \label{eq:coupled_kin}
        & -e\vec{E}\vec{v}_{{\bm k}\uparrow} \frac{\partial f^0_{{\bm k}\uparrow}}{\partial \varepsilon_{{\bm k}\uparrow}} =-\frac{\delta f_{{\bm k}\uparrow}}{\tau_{\rm tr}}-\frac{\delta f_{{\bm k}\uparrow}-\delta f_{{\bm k}\downarrow}}{\tau_{\rm sf}} \\
        & -e\vec{E}\vec{v}_{{\bm k}\downarrow} \frac{\partial f^0_{{\bm k}\downarrow}}{\partial \varepsilon_{{\bm k}\downarrow}} =-\frac{\delta f_{{\bm k}\downarrow}}{\tau_{\rm tr}}-\frac{\delta f_{{\bm k}\downarrow}-\delta f_{{\bm k}\uparrow}}{\tau_{\rm sf}}
    \end{aligned}
    \right.
\end{equation}
{To solve them,}
we now introduce linear combinations of distribution functions
\begin{align}
    & f^0_{{\bm k}+} = f^0_{{\bm k}\uparrow}+f^0_{{\bm k}\downarrow} \qquad f^0_{{\bm k}-} = f^0_{{\bm k}\uparrow}-f^0_{{\bm k}\downarrow} \label{eq:inversetransform1}\\
    & \delta f_{{\bm k}+} = \delta f_{{\bm k}\uparrow}+\delta f_{{\bm k}\downarrow} \qquad \delta f_{{\bm k}-} = \delta f_{{\bm k}\uparrow} - \delta f_{{\bm k}\downarrow} \label{eq:inversetransform2}\\
    & \delta f_{{\bm k}\uparrow} = \frac{1}{2}\left( \delta f_{{\bm k}+}+\delta f_{{\bm k}-} \right) \qquad\delta f_{{\bm k}\downarrow} = \frac{1}{2}\left( \delta f_{{\bm k}+}-\delta f_{{\bm k}-} \right) 
\end{align}
In terms of this new variables the equations{~(\ref{eq:coupled_kin})} decouple and we have
\begin{equation}
    \left\{ 
    \begin{aligned} \label{eq:coupled_kin2}
        & e\vec{E}\vec{v}_{{\bm k}} \frac{\partial f^0_{{\bm k}+}}{\partial \mu} =-\frac{\delta f_{{\bm k}+}}{\tau_{\rm tr}} \\
        & e\vec{E}\vec{v}_{{\bm k}} \frac{\partial f^0_{{\bm k}-}}{\partial \mu} =-2\frac{\delta f_{{\bm k}-}}{\tau_{\rm sf}}
    \end{aligned}
    \right.
\end{equation}
Therefore, we find the expressions for $\delta f_{\vec{k}+(-)}$. Now our task is to plug the inverse transformations ~(\ref{eq:inversetransform1})-(\ref{eq:inversetransform2}) into the microscopically derived collision integral \eqref{linspinup}-\eqref{linspindown} and explicitly express $\tau^{-1}_{\rm tr}$ and $\tau^{-1}_{\rm sf}$. 
Eq.~\eqref{linspinup} becomes
\begin{align}
     & \frac{J^2}{\pi} \sum_{{\bm p}} \int \limits_{-\infty}^\infty d \omega \delta\left( \varepsilon_{{\bm k}\uparrow}-\varepsilon_{{\bm p}\downarrow}-\omega \right) \left( \Im \chi^x_{{\bm k}-p} (\omega)+\Im \chi^y_{{\bm k}-p} (\omega) \right) (TD)_{\uparrow \downarrow}\left[ \frac{\frac{1}{2}\left( \delta f_{{\bm k}+}+\delta f_{{\bm k}-} \right)}{(1-f_{{\bm k} \uparrow}^0)^2} - \frac{\frac{1}{2}\left( \delta f_{{\bm p}+}-\delta f_{{\bm p}-} \right)}{(1-f^0_{{\bm p} \downarrow})^2}\frac{n_B(\omega)}{n_B(\omega)+1} \right] + \nonumber \\
     & \frac{J^2}{\pi} \sum_{{\bm p}} \int \limits_{-\infty}^\infty d \omega \delta\left( \varepsilon_{{\bm k}\uparrow}-\varepsilon_{{\bm p}\uparrow}-\omega \right) \Im \chi^z_{{\bm k}-p} (\omega) (TD)_{\uparrow \uparrow}\left[ \frac{\frac{1}{2}\left( \delta f_{{\bm k}+}+\delta f_{{\bm k}-} \right)}{(1-f_{{\bm k} \uparrow}^0)^2} - \frac{\frac{1}{2}\left( \delta f_{{\bm p}+}+\delta f_{{\bm p}-} \right)}{(1-f^0_{{\bm p} \uparrow})^2}\frac{n_B(\omega)}{n_B(\omega)+1} \right]
\end{align}
This expression, together with the LHS of the kinetic equation{~(\ref{eq:coupled_kin2})}, is to be multiplied by $\cos \phi_k$ and integrated over $d^2{\bm k}$. To 
{lighten the notation, we introduce the measure}
% save some space and time let us introduce the measure 
\begin{equation}
    \int d\mu_{\sigma \lambda} = \frac{J^2}{2\pi} \int \limits_0^{2\pi}\frac{d\phi_k}{(2\pi)^2}\int\limits_0^{2\pi}\frac{d\phi_p}{(2\pi)^2}\int\limits_{-\infty}^\infty d\omega \int\limits_0^\infty (dk dp) kp \cos \phi_k \delta \left( \varepsilon_{{\bm k}\sigma}-\varepsilon_{{\bm p}\lambda}-\omega \right)
\end{equation}
The kinetic equation for the up-spin distribution function can then be written as
% in the following way 
\begin{align}
    & \int\frac{dk d\phi_k}{(2\pi)^2} e E v_{0}k\cos^2\phi_k \frac{\partial f^0_{{\bm k}\uparrow}}{\partial \mu} = \nonumber \\
    & \int d\mu_{\uparrow \downarrow} \left( \Im \chi^x_{{\bm k}-p} (\omega)+\Im \chi^y_{{\bm k}-p} (\omega) \right) (TD)_{\uparrow \downarrow}\left[ \frac{\delta f_{{\bm k}+}}{(1-f_{{\bm k} \uparrow}^0)^2} - \frac{\delta f_{{\bm p}+}}{(1-f^0_{{\bm p} \downarrow})^2}\frac{n_B(\omega)}{n_B(\omega)+1} \right] +\nonumber\\
    &\int d\mu_{\uparrow \uparrow} \Im \chi^z_{{\bm k}-p} (\omega) (TD)_{\uparrow \uparrow}\left[ \frac{\delta f_{{\bm k}+}}{(1-f_{{\bm k} \uparrow}^0)^2} - \frac{\delta f_{{\bm p}+}}{(1-f^0_{{\bm p} \uparrow})^2}\frac{n_B(\omega)}{n_B(\omega)+1} \right] +\nonumber\\
    & \int d\mu_{\uparrow \downarrow} \left( \Im \chi^x_{{\bm k}-p} (\omega)+\Im \chi^y_{{\bm k}-p} (\omega) \right) (TD)_{\uparrow \downarrow}\left[ \frac{\delta f_{{\bm k}-}}{(1-f_{{\bm k} \uparrow}^0)^2} + \frac{\delta f_{{\bm p}-}}{(1-f^0_{{\bm p} \downarrow})^2}\frac{n_B(\omega)}{n_B(\omega)+1} \right] +\nonumber\\
    &\int d\mu_{\uparrow \uparrow} \Im \chi^z_{{\bm k}-p} (\omega) (TD)_{\uparrow \uparrow}\left[ \frac{\delta f_{{\bm k}-}}{(1-f_{{\bm k} \uparrow}^0)^2} - \frac{\delta f_{{\bm p}-}}{(1-f^0_{{\bm p} \uparrow})^2}\frac{n_B(\omega)}{n_B(\omega)+1} \right]
\end{align}
where the first two lines of the RHS are altogether proportional to $\tau_{\rm tr}$ and the third and fourth combined are proportional to $\tau_{\rm sf}$. We can write this equation as
\begin{equation}
     \left\{ 
    \begin{aligned}
        & eEv_0\frac{\partial n_{0\uparrow}}{\partial \mu} = A \tau_{\rm tr} + B\tau_{\rm sf} \\
        & eEv_0\frac{\partial n_{0\downarrow}}{\partial \mu} = C \tau_{\rm tr} + D\tau_{\rm sf} 
    \end{aligned}
    \right.
    ,
\end{equation}
where we have introduced 
\begin{align}
    A &= \int d\mu_{\uparrow \downarrow} \left( \Im \chi^x_{{\bm k}-p} (\omega)+\Im \chi^y_{{\bm k}-p} (\omega) \right) (TD)_{\uparrow \downarrow}\left[ \frac{ - \cos \phi_k \partial_\mu f_{{\bm k}+}^0 }{(1-f_{{\bm k} \uparrow}^0)^2} - \frac{- \cos \phi_p \partial_\mu f_{{\bm p}+}^0}{(1-f^0_{{\bm p} \downarrow})^2}\frac{n_B(\omega)}{n_B(\omega)+1} \right]
    \nonumber\\
    &+
    \int d\mu_{\uparrow \uparrow} \Im \chi^z_{{\bm k}-p} (\omega) (TD)_{\uparrow \uparrow}\left[ \frac{- \cos \phi_k \partial_\mu f_{{\bm k}+}^0}{(1-f_{{\bm k} \uparrow}^0)^2} - \frac{- \cos \phi_p \partial_\mu f_{{\bm p}+}^0}{(1-f^0_{{\bm p} \uparrow})^2}\frac{n_B(\omega)}{n_B(\omega)+1} \right] ,
    \\
    B &= \frac{1}{2} \int d\mu_{\uparrow \downarrow} \left( \Im \chi^x_{{\bm k}-p} (\omega)+\Im \chi^y_{{\bm k}-p} (\omega) \right) (TD)_{\uparrow \downarrow}\left[ \frac{ \cos \phi_k \partial_\mu f_{{\bm k}-}^0 }{(1-f_{{\bm k} \uparrow}^0)^2} + \frac{\cos \phi_p \partial_\mu f_{{\bm p}-}^0}{(1-f^0_{{\bm p} \downarrow})^2}\frac{n_B(\omega)}{n_B(\omega)+1} \right]
    \nonumber\\
    &+
    \frac{1}{2}\int d\mu_{\uparrow \uparrow} \Im \chi^z_{{\bm k}-p} (\omega) (TD)_{\uparrow \uparrow}\left[ \frac{\cos \phi_k \partial_\mu f_{{\bm k}-}^0}{(1-f_{{\bm k} \uparrow}^0)^2} - \frac{\cos \phi_p \partial_\mu f_{{\bm p}-}^0}{(1-f^0_{{\bm p} \uparrow})^2}\frac{n_B(\omega)}{n_B(\omega)+1} \right] ,
    \\
    C &= \int d\mu_{\downarrow \uparrow} \left( \Im \chi^x_{{\bm k}-p} (\omega)+\Im \chi^y_{{\bm k}-p} (\omega) \right) (TD)_{\downarrow \uparrow}\left[ \frac{ - \cos \phi_k \partial_\mu f_{{\bm k}+}^0 }{(1-f_{{\bm k} \downarrow}^0)^2} - \frac{- \cos \phi_p \partial_\mu f_{{\bm p}+}^0}{(1-f^0_{{\bm p} \uparrow})^2}\frac{n_B(\omega)}{n_B(\omega)+1} \right]
    \nonumber\\
    &+ 
    \int d\mu_{\downarrow \downarrow} \Im \chi^z_{{\bm k}-p} (\omega) (TD)_{\downarrow \downarrow}\left[ \frac{- \cos \phi_k \partial_\mu f_{{\bm k}+}^0}{(1-f_{{\bm k} \downarrow}^0)^2} - \frac{- \cos \phi_p \partial_\mu f_{{\bm p}+}^0}{(1-f^0_{{\bm p} \downarrow})^2}\frac{n_B(\omega)}{n_B(\omega)+1} \right] ,
    \\
    D &= -\frac{1}{2} \int d\mu_{\downarrow \uparrow} \left( \Im \chi^x_{{\bm k}-p} (\omega)+\Im \chi^y_{{\bm k}-p} (\omega) \right) (TD)_{\downarrow \uparrow}\left[ \frac{ \cos \phi_k \partial_\mu f_{{\bm k}-}^0 }{(1-f_{{\bm k} \downarrow}^0)^2} + \frac{\cos \phi_p \partial_\mu f_{{\bm p}-}^0}{(1-f^0_{{\bm p} \uparrow})^2}\frac{n_B(\omega)}{n_B(\omega)+1} \right]
    \nonumber\\
    &-\frac{1}{2}\int d\mu_{\downarrow \downarrow} \Im \chi^z_{{\bm k}-p} (\omega) (TD)_{\downarrow \downarrow}\left[ \frac{\cos \phi_k \partial_\mu f_{{\bm k}-}^0}{(1-f_{{\bm k} \downarrow}^0)^2} - \frac{\cos \phi_p \partial_\mu f_{{\bm p}-}^0}{(1-f^0_{{\bm p} \downarrow})^2}\frac{n_B(\omega)}{n_B(\omega)+1} \right].
\end{align}
% We can formally 
{This allows us to}
write the explicit solution for the relaxation times
\begin{equation}
     \left\{ 
    \begin{aligned}
        & \tau_{\rm tr} = \frac{eEv_0}{AD-BC}\left( D \frac{\partial n_{0\uparrow}}{\partial \mu} - B \frac{\partial n_{0\downarrow}}{\partial \mu} \right),\\
        & \tau_{\rm sf} = \frac{eEv_0}{AD-BC}\left( A \frac{\partial n_{0\downarrow}}{\partial \mu} - C \frac{\partial n_{0\uparrow}}{\partial \mu} \right) .
    \end{aligned}
    \right.
\end{equation}

\section{Effect of magnetic field on transport}\label{app:mag_field}
The Kitaev spin Hamiltonian \cite{kitaev2006anyons} in an applied magnetic field is 
\be
    H_m = -J_K\sum_{\langle i,j \rangle_\gamma} s_i^\gamma s_j^\gamma + \sum_{i} \vec{h}\cdot \vec{s}_i.
\ee
This effective model describes magnetic moments located at the Ru sites of $\alpha$-RuCl$_3$ for temperatures above 7 K \cite{Banerjee2017_Science_RuCl3_QSL, PhysRevLett.102.017205}. This model can only be exactly solved for $\vec{h} = 0$ \cite{kitaev2006anyons}. However, for small applied magnetic fields ($h\ll J_K$) the sector of low-energy excitations can be described by an effective hamiltonian 
\begin{equation}
    H_g = -J_K\sum_{\langle i,j \rangle_\gamma} s_i^\gamma s_j^\gamma - g \sum_{\langle \langle i,k \rangle \rangle} s_i^\gamma s_j^{\gamma'} s_k^{\gamma''} ,
\end{equation}
where $g = h_xh_yh_z/\Delta^2$ and $\langle \langle i,k \rangle \rangle$ is a pair of next-to-nearest neighbors such that $j$ is the only site connected to both $i$ and $k$. 
\begin{figure}[h]
\centering
    \begin{overpic}[width = 0.45\linewidth]{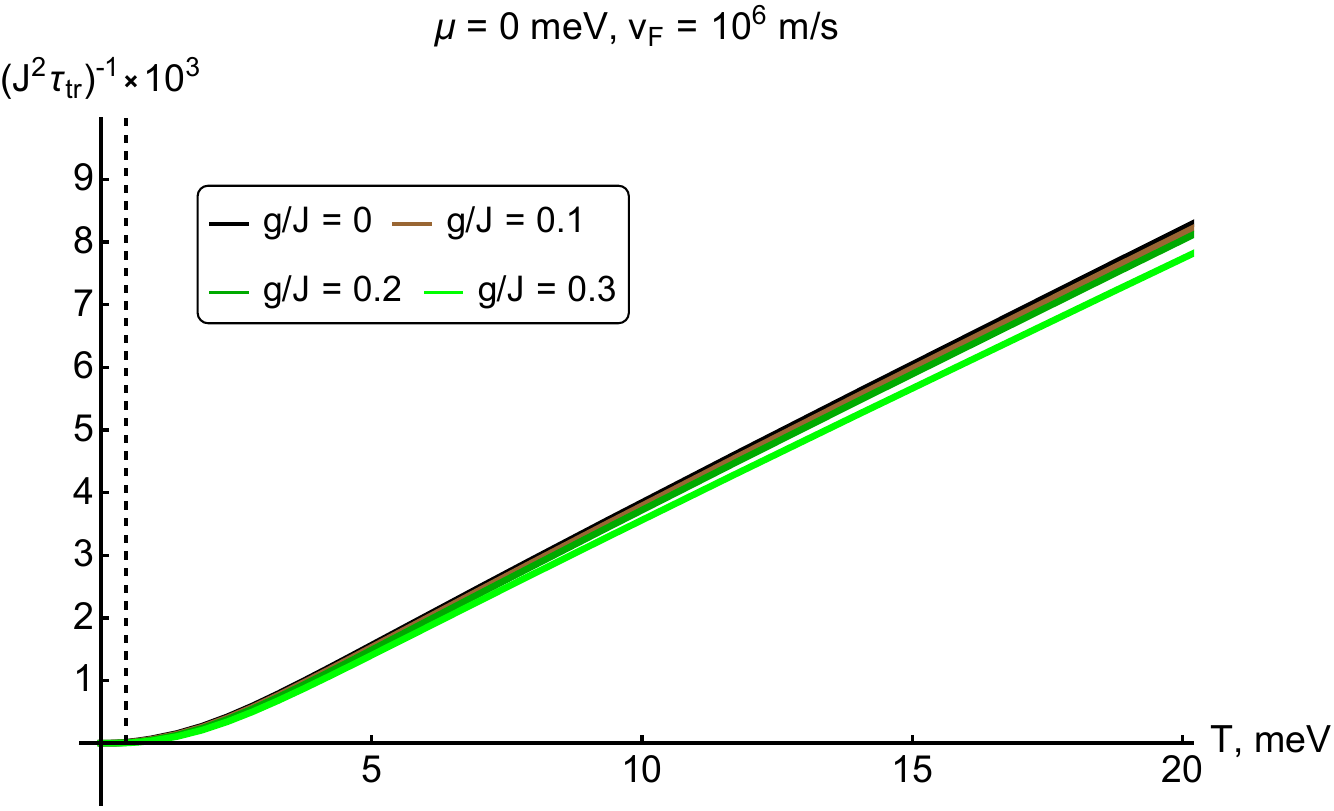}
    \put(210,20){{\large (a)}} 
    \end{overpic} \quad 
    \begin{overpic}[width = 0.45\linewidth]{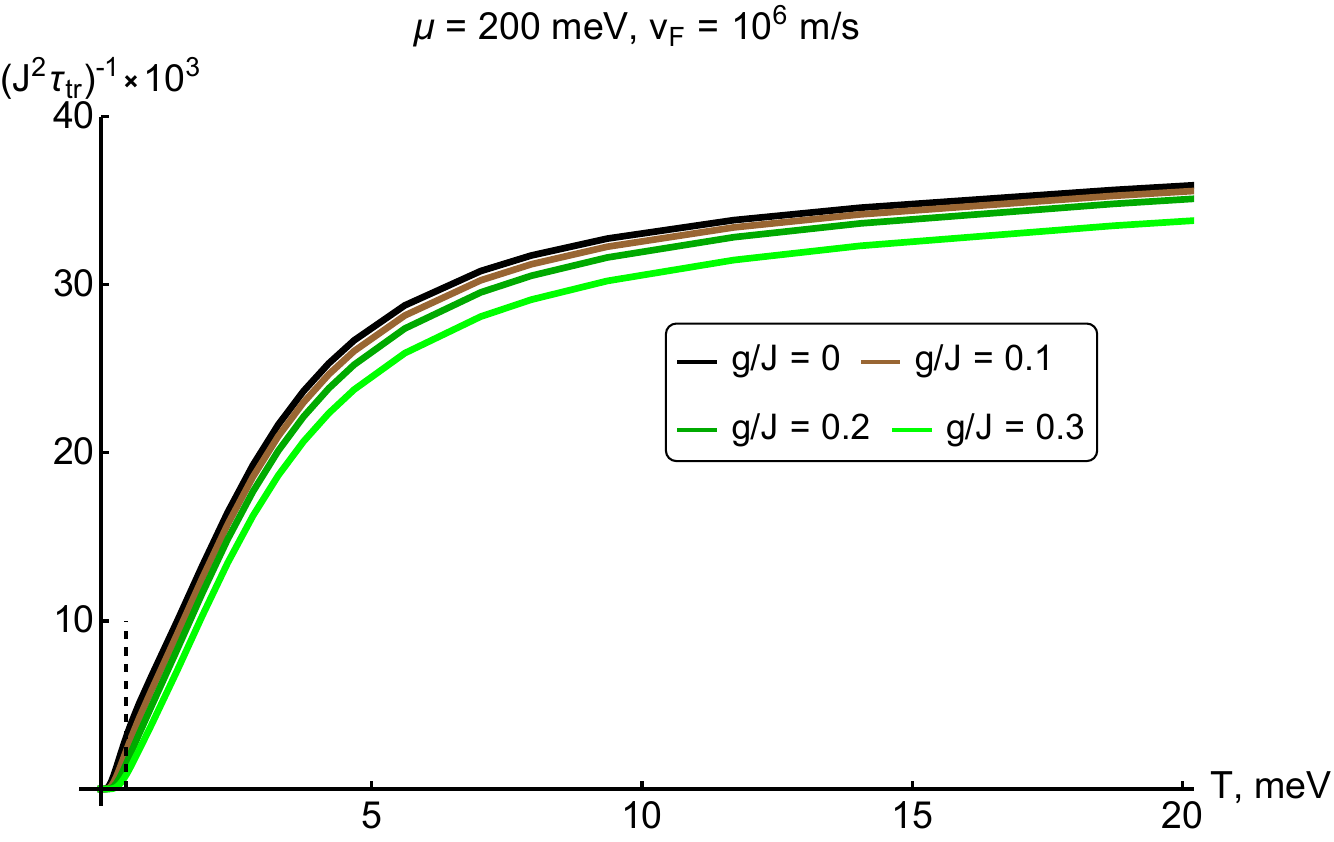}
    \put(210,20){{\large (b)}} 
    \end{overpic}
\caption{\label{fig:mag_field}  
Panel (a):
The solid lines are normalized collision rate $(\tau_{\rm tr} J^2)^{-1}$, Eq.~\eqref{tausf}, for different values of external magnetic field as a function of temperature $T$ for a Fermi velocity $v_{\rm F} = 10^6~{\rm m/s}$. The chemical potential is $\mu = 0$ eV. 
Panel (b):
Same as Panel (a), but for $\mu = 200$ meV.
}
\end{figure}
We calculate numerically the Green's functions $\Im Q_c^{1,\gamma}(\omega)$ in presence of applied magnetic field as discussed in Ref.~\cite{carrega2020tunneling}. The response function $\Im \chi(\vec{k},\omega)$ depends on the magnetic field via $\Im Q_c^{1,\gamma}(\omega)$ and $\Delta$ which is also calculated numerically for different values of magnetic field~\cite{carrega2020tunneling}, expressed as a dimensionless parameter $g/J_K$. We use these quantities to calculate the transport time $\tau_{\rm tr}$. \\
The transport rates are presented on Fig.~\ref{fig:mag_field} for different dopoings. We report very low effect of magnetic field on spin-unpolarised charge currents. The highest ratio $$\frac{\tau_{\rm tr}({g/J_K=0})}{\tau_{\rm tr}({g/J_K=0.3})}\approx3.6$$ is achieved for $\mu=200$ meV at $k_{\rm B}T=\Delta(g/J_K=0)$. The relative difference becomes smaller at both high temperatures ($k_{\rm B}T\gg\Delta(g/J_K=0)$) and small ones ($k_{\rm B}T\ll\Delta(g/J_K=0)$). 

\end{widetext}

\end{document}

%% file: main.bbl
%apsrev4-2.bst 2019-01-14 (MD) hand-edited version of apsrev4-1.bst
%Control: key (0)
%Control: author (8) initials jnrlst
%Control: editor formatted (1) identically to author
%Control: production of article title (0) allowed
%Control: page (0) single
%Control: year (1) truncated
%Control: production of eprint (0) enabled
%